\RequirePackage{fix-cm}
\documentclass[smallcondensed]{svjour3}
\smartqed  
\usepackage{cite}
\usepackage{hyperref}
\usepackage[numbered]{bookmark}
\usepackage{soul}
\usepackage{booktabs}
\usepackage{multirow}
\usepackage{float}
\usepackage{url}
\usepackage[small,it]{caption}
\usepackage{tcolorbox}
\usepackage{amsmath,amssymb,amsfonts}
\usepackage{algorithmic}
\usepackage{graphicx}
\usepackage{textcomp}
\usepackage{xcolor}
\usepackage{tabularx} 
\usepackage{balance}
\usepackage{enumitem}
\usepackage[a-2b,mathxmp]{pdfx}[2018/12/22]

\usepackage[utf8]{inputenc}
\usepackage{newunicodechar,graphicx}
\DeclareRobustCommand{\okina}{%
  \raisebox{\dimexpr\fontcharht\font`A-\height}{%
    \scalebox{0.8}{`}%
  }%
}
\newunicodechar{ʻ}{\okina}

\journalname{Empirical Software Engineering}

\newcommand{\RQA}{What behavioral roles do closed-category terms play in source code identifiers?}

\newcommand{\RQB}{How do closed-category terms correlate with structural, programming language, and domain-specific contexts in software?}

\begin{document}

\title{On the Structure and Semantics of Identifier Names Containing Closed Syntactic Category Words}

\author{Christian D. Newman     \and
Anthony Peruma       \and
Eman Abdullah AlOmar          \and
Mahie Crabbe         \and
Syreen Banabilah     \and
Reem S. Alsuhaibani  \and
Michael J. Decker    \and
Farhad Akhbardeh     \and
Marcos Zampieri      \and
Mohamed Wiem Mkaouer \and
Jonathan I. Maletic
}

\institute{ Christian D. Newman \at
            \email{cdnvse@rit.edu}
            \and
            Anthony Peruma \at
            \email{peruma@hawaii.edu}
            \and
            Eman Abdullah AlOmar \at
            \email{ealomar@stevens.edu}
            \and
            Mahie Crabbe \at
            \email{mahi3@hawaii.edu}
            \and
            Syreen Banabilah \at
            \email{sbanabil@kent.edu}
            \and
            Reem S. AlSuhaibani \at
            \email{rsuhaibani@psu.edu.sa}
            \and
            Michael J. Decker \at
            \email{mdecke@bgsu.edu}
            \and
            Farhad Akhbardeh \at
            \email{farhad.akhbardeh@briarcliff.edu}
            \and
            Marcos Zampieri \at
            \email{mzampier@gmu.edu}
            \and
            Mohamed Wiem Mkaouer \at
            \email{mmkaouer@umich.edu}
            \and
            Jonathan I. Maletic \at
            \email{jmaletic@kent.edu}
}


 \maketitle

\begin{abstract}
Identifier names are crucial components of code, serving as primary clues for developers to understand program behavior. This paper investigates the linguistic structure of identifier names by extending the concept of grammar patterns, which represent the part-of-speech (PoS) sequences underlying identifier phrases. The specific focus is on closed syntactic categories (e.g., prepositions, conjunctions, determiners), which are rarely studied in software engineering despite their central role in general natural language. To study these categories, the Closed Category Identifier Dataset (CCID), a new manually annotated dataset of 1,275 identifiers drawn from 30 open-source systems, is constructed and presented. The relationship between closed-category grammar patterns and program behavior is then analyzed using grounded-theory-inspired coding, statistical, and pattern analysis. The results reveal recurring structures that developers use to express concepts such as control flow, data transformation, temporal reasoning, and other behavioral roles through naming. This work contributes an empirical foundation for understanding how linguistic resources encode behavior in identifier names and supports new directions for research in naming, program comprehension, and education.

\keywords{identifier naming \and program comprehension \and part of speech tagging \and software maintenance and evolution \and software linguistics \and closed category terms \and naming conventions}
\end{abstract}

\section{Introduction}
\label{introduction}

Developers spend a significant amount of time reading and comprehending code \cite{Corbi1989,Martin:2008}, and identifier names play a central role in this process, accounting for roughly 70\% of all code characters \cite{Deissenbock:2005}. Prior work shows that the quality of identifier names significantly impacts comprehension \cite{Schankin:2018,Binkley2006,Hofmeister:2017,butler2010exploring,Takang1996,feitelson:2017,fakhoury:2020}, supports tooling \cite{Binkley:2018, newmanabbrev}, and poses persistent pedagogical challenges \cite{Felienne:2024, Glassman2015FoobazVN}. These challenges motivate research into how naming practices encode meaning, and how we might better characterize or improve them.

A key obstacle in studying identifier names is measuring the semantics they convey, not just at the level of individual terms, but in the structure and composition of entire names. Some approaches cluster identifiers by terms or embeddings \cite{Allamanis:2015, Liu:2019}, while others analyze them using syntactic or static roles \cite{Dragan:2006, Alsuhaibani:2015, Newman:2017}. In this work, we focus instead on \textbf{grammar patterns}\cite{Newman_GP}: sequences of part-of-speech (PoS) tags that abstract the phrasal structure of identifiers. Grammar patterns provide a syntactic lens through which naming semantics can be studied at scale, offering insight into how term combinations convey behavioral meaning.

At a high level, PoS can be split into two Syntactic Categories: \textbf{open} and \textbf{closed}. Most identifier naming research has focused on \textbf{open-category}, which includes nouns and verbs. The set of open category terms evolves and expands (in terms of new words) over time as new domains emerge and evolve. In contrast, \textbf{closed-category} (e.g., prepositions, conjunctions, determiners) are drawn from a fixed set and serve functional roles in language; this set of terms rarely sees new words introduced over time. These terms have received little attention in the software literature, despite their importance in human languages. Identifying closed-category terms in code is also nontrivial: for example, the word \texttt{and} may represent a conjunction or a logical operator, depending on context, making PoS tagging a prerequisite for meaningful analysis.

The goal of this paper is to investigate how \textbf{closed-category terms} are used in identifier names to express program behavior, using the grammar patterns (see Section~\ref{grammarpatterndef} for definitions) that these terms appear within to provide insights into how these terms interact with the other terms around them. We extend prior research on general grammar patterns~\cite{Newman_GP, newman2021Catalogue} by introducing and analyzing the Closed Category Identifier Dataset (CCID), a manually annotated corpus of 1,275 identifiers from 30 open-source systems. Unlike raw term-based approaches, grammar patterns abstract away surface vocabulary, allowing us to characterize naming conventions by their syntactic structure. By examining both the patterns and the concrete terms that instantiate them, we explore how developers use compact linguistic forms to encode behavioral semantics in code. Specifically, we contribute:

\begin{itemize}
    \item \textbf{A new dataset (CCID)} of identifiers containing closed-category terms, annotated with PoS tags, grammar patterns, and contextual metadata.
    \item \textbf{A mixed-methods analysis} combining grounded-theory-style coding with statistical evaluation to characterize the semantics of closed-category grammar patterns and their constituent terms.
    \item \textbf{An evaluation of how these patterns correlate with programming context, language, and domain}.
\end{itemize}

Our findings have implications for both human and automated naming support. Grammar patterns provide a structured approach to analyzing naming behavior, identifying potential inconsistencies, and providing naming suggestions. For AI-based tools, they offer scaffolding to align generated names with human conventions. For developers and educators, they reveal naming idioms that can support clearer communication and pedagogy. In this study, we address the following research questions:

RQ1: \textbf{\RQA} To address this question, we conducted a grounded-theory-inspired study on a manually annotated dataset of identifiers containing closed-category terms: prepositions, conjunctions, determiners, and numerals. Through open coding and memoing, we develop axial and selective codes that describe the behavioral functions these terms convey in source code, such as data flow, condition handling, or execution sequencing. This process allows us to uncover not only common grammar patterns but also the communicative intent behind developers’ use of closed-category terms. Our goal is to characterize the nuanced and purposeful ways in which these terms encode program behavior and convey information.

RQ2: \textbf{\RQB} To answer this question, we quantitatively analyze the distribution of closed-category terms across multiple dimensions: source-code-local structure (e.g., function names, parameters, class names), programming languages (e.g., Java, C++, C), and system domains (e.g., libraries, frameworks, domain-specific applications). We use statistical tests to examine whether these terms appear disproportionately in specific contexts. These correlations help us determine whether developers systematically leverage closed-category terms to express behavior in ways that are shaped by structural conventions, linguistic norms, or domain constraints.

This paper is organized as follows.  Section \ref{motivation} provides our reasoning on why it is essential to study this topic.  Section \ref{grammarpatterndef} gives background on grammar patterns in the context of identifier names.  Section \ref{methodology} provides a detailed explanation of the methods used for conducting the investigation. Our Evaluations are presented in Sections \ref{rqadiscussion} and \ref{rqbdiscussion}.  Related work on identifier names is in Section \ref{sec:related}.   Discussion of the results is in section \ref{discussion}, followed by Threats to Validity in Section \ref{threats}.  Conclusions are in Section \ref{conclusions} and Data Availability in Section \ref{data_avail}.


\section{Why Study Closed-Category Naming Patterns?}\label{motivation}

Closed-category terms are relatively uncommon in identifier names. Because they are uncommon, their presence raises an important question: \textit{When developers do use these terms, what specific meaning or behavior are they trying to convey?} We hypothesize that developers include closed-category terms deliberately, as a way to encode behaviorally specific semantics that are lost or obscured without them. Consider the following examples:

\begin{itemize}
    \item \texttt{find\_all\_textures}: The determiner \texttt{all} signals a universal scope, clarifying that this identifier refers to the entire set of textures, not a subset.
    \item \texttt{on\_start}: The preposition \texttt{on} reflects event-driven logic, indicating that the associated behavior is triggered at the start of execution.
    \item \texttt{warn\_if\_error}: The conjunction \texttt{if} embeds a conditional relationship, revealing that the action is contingent on an error occurring.
\end{itemize}

In each case, the closed-category term is essential to understanding the behavioral semantics of the identifier. Without these terms, the names are more ambiguous or less informative. While uncommon in aggregate, closed-category terms often signal precise intent and encode logical structure in compact forms.

Despite their potential significance, these terms have received almost no attention in prior software development naming research, which has focused primarily on open-category words (e.g., nouns, verbs). As a result, we lack foundational knowledge about when and how closed-category terms are used in code and what they contribute to program comprehension.

Understanding these naming patterns has clear implications: it can inform naming tools, guide educational resources, improve automated name generation, and help researchers characterize naming conventions more precisely. Closed-category terms may be uncommon, but we argue, in this paper, that their usage is not accidental; they significantly contribute to the meaning of identifier names, making it important to study them. More examples of identifiers containing closed-category terms can be found in Table \ref{tab:closed-category-grammar-patterns}.

\begin{table}[]
\centering
\caption{Examples of closed-category grammar patterns}
\label{tab:closed-category-grammar-patterns}
\begin{tabular}{@{}ll@{}}
\toprule
Identifier Example & Grammar Pattern \\ \midrule
action to index map & N P NM N \\
as binary & P N \\
time for each line & N P DT N \\
server and port & N CJ N \\
open if empty & V CJ NM \\
adjust to camera & V P N \\ \bottomrule
\end{tabular}%
\end{table}
\begin{table}[]
\setlength{\tabcolsep}{1.5pt}
\centering
\caption{Part-of-speech categories used in study}
\label{tab:posusedtable}
\begin{tabular}{@{}lll@{}}
\toprule
\textbf{Abbreviation} & \textbf{Expanded Form} & \textbf{Examples} \\ \midrule
N & noun & \begin{tabular}[c]{@{}l@{}}stack, function, language\end{tabular} \\ \midrule
DT & determiner & \begin{tabular}[c]{@{}l@{}}the, this, that, these, those, which\end{tabular} \\ \midrule
CJ & conjunction & and, for, nor, but, or, yet, so \\ \midrule
P & preposition & \begin{tabular}[c]{@{}l@{}}behind, in front of, at, under,\\ beside, above, beneath, despite\end{tabular} \\ \midrule
NPL & noun plural & \begin{tabular}[c]{@{}l@{}}strings, identifiers, classes\end{tabular} \\ \midrule
NM (bold) & noun modifier & \begin{tabular}[c]{@{}l@{}}\textbf{employee}Name, \textbf{token}Parser\end{tabular} \\ \midrule
V & verb & run, execute, implement, develop \\ \midrule
VM & verb modifier (adverb) & \begin{tabular}[c]{@{}l@{}}quickly, safely, eventually\end{tabular} \\ \midrule
PR & pronoun & \begin{tabular}[c]{@{}l@{}}she, he, her, him, it, we,\\ us, they, them, I, me, you\end{tabular} \\ \midrule
D & numeral & 1, 2, 10, 4.12, 0xAF \\ \midrule
PRE & preamble* & Gimp, GLEW, GL, G \\ \bottomrule
\end{tabular}
\end{table}

\begin{figure}[h]
\centering
\includegraphics{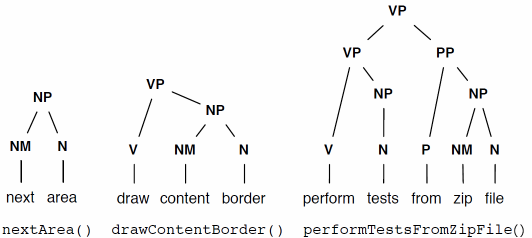}
\caption{Examples of noun, verb, and prepositional phrases}
\label{fig:example_phrasal_concepts}
\end{figure}

\section{Definitions \& Grammar Pattern Generation}
\label{grammarpatterndef}

In this work, we analyze identifier names through the lens of \textbf{grammar patterns}, which are sequences of part-of-speech (PoS) tags assigned to the terms within an identifier. For example, the identifier \texttt{GetUserToken} is split into the terms \texttt{Get}, \texttt{User}, and \texttt{Token}, which are tagged as \texttt{Verb Noun-adjunct Noun}. This sequence, \texttt{V NM N}, represents the identifier’s grammar pattern. Crucially, this pattern generalizes across many identifiers: \texttt{RunUserQuery} and \texttt{WriteAccessToken} share the same structure, despite using different terms. Grammar patterns thus allow us to relate identifiers by their syntactic form.

We focus specifically on \textbf{closed-category grammar patterns}, which are patterns that contain at least one closed-class part of speech: a \textbf{preposition}, \textbf{determiner}, \textbf{conjunction}, or \textbf{numeral}. These categories are finite and rarely accept new terms, in contrast to open-class categories like nouns and verbs, which grow over time as new domains introduce new concepts. Despite their rarity in code, closed-category terms often signal behavioral relationships such as event triggers, quantification, or conditional logic, making them critical to study.

\textbf{Part-of-Speech Tags.} Table~\ref{tab:posusedtable} lists the PoS tags used in this study. Most are drawn from standard linguistic categories. We highlight one custom tag that is central to our analysis:

\begin{itemize}
    \item \textbf{Noun Modifier (NM):} Includes adjectives as well as \textit{noun-adjuncts}—nouns used to modify another noun (e.g., \texttt{user} in \texttt{userToken}, or \texttt{content} in \texttt{contentBorder}). Although standard PoS taggers do not typically distinguish noun-adjuncts, prior work shows their critical role in naming semantics \cite{Newman_GP}.
    
    \item \textbf{Preamble (PRE):} A prefix used to convey structural or language-specific metadata, rather than domain semantics. Common examples include Hungarian-style markers such as \texttt{m\_} for member variables, or project-level namespaces like \texttt{gimp} in \texttt{gimp\_temp\_file}; a practice especially common in C. For a complete typology and discussion, see \cite{Newman_GP}; we include preambles here since we do use them in the data set, but they are not the focus of this paper.

\end{itemize}

\subsection{Phrasal Structures and Interpretation}
\label{phrasalconcepts}

While our analysis is based on PoS sequences rather than full parse trees, we draw on linguistic phrase structure to interpret identifier patterns. Specifically, we reference three example concepts to help the reader understand what we mean when we use the term `phrase' with respect to grammar patterns:

\begin{itemize}
    \item \textbf{Noun Phrase (NP):} A noun optionally preceded by one or more modifiers (e.g., \texttt{accessLog}, \texttt{userToken}, \texttt{windowTitle}).
    \item \textbf{Verb Phrase (VP):} A verb followed by a noun phrase, often representing an action on a specific entity (e.g., \texttt{getUserToken}, \texttt{drawContentBorder}).
    \item \textbf{Prepositional Phrase (PP):} A preposition followed by a noun phrase (e.g., \texttt{onClick}, \texttt{fromCache}).
\end{itemize}

These phrase structures help illustrate how grammar patterns support analysis of phrases. For instance, in \texttt{drawContentBorder}, the noun-modifier \texttt{content} refines the meaning of the head noun \texttt{border}, while the verb \texttt{draw} anchors the identifier as a behavior applied to that concept (i.e., draw applied to a specific type of border; a content-border). When closed-category terms appear, they may indicate when an action should occur (\texttt{onStart}), under what condition (\texttt{ifError}), or which entities are included (\texttt{allTextures}). Figure~\ref{fig:example_phrasal_concepts} shows examples of NP, VP, and VP-with-PP constructions as derived from grammar patterns.

\begin{table}[]
\centering
\caption{List of 30 open source systems included in study}
\label{tab:study_oss}
\resizebox{\textwidth}{!}{%
\begin{tabular}{@{}llllllll@{}}
\toprule
Repo Link                                         & Name           & Primary Language & Data of most recent commit & C LOC   & C++ LOC & Java LOC & Total LOC \\ \midrule
https://github.com/liuliu/ccv                     & ccv            & C                & 2023-07-19                 & 279186  & 1908    & 0        & 281094    \\ \midrule
https://github.com/ropensci/git2r                 & git2r          & C                & 2023-05-01                 & 99956   & 0       & 0        & 99956     \\ \midrule
https://github.com/Juniper/libxo                  & libxo          & C                & 2023-02-08                 & 9361    & 0       & 0        & 9361      \\ \midrule
https://github.com/mgba-emu/mgba                  & mgba           & C                & 2023-07-18                 & 302865  & 25309   & 0        & 328174    \\ \midrule
https://github.com/naemon/naemon-core             & naemon-core    & C                & 2023-07-07                 & 40991   & 0       & 0        & 40991     \\ \midrule
https://github.com/openvswitch/ovs                & ovs            & C                & 2023-07-19                 & 268376  & 0       & 0        & 268376    \\ \midrule
https://github.com/igraph/rigraph                 & rigraph        & C                & 2023-07-19                 & 273973  & 32737   & 0        & 306710    \\ \midrule
https://github.com/toggl-open-source/toggldesktop & toggldesktop   & C                & 2023-06-22                 & 582087  & 269105  & 0        & 851192    \\ \midrule
https://github.com/irungentoo/toxcore             & toxcore        & C                & 2018-10-03                 & 27443   & 0       & 0        & 27443     \\ \midrule
https://github.com/weechat/weechat                & weechat        & C                & 2023-07-20                 & 197608  & 31175   & 0        & 228783    \\ \midrule
https://github.com/wireshark/wireshark            & wireshark      & C                & 2023-07-20                 & 4171790 & 102848  & 0        & 4274638   \\ \midrule
https://github.com/BVLC/caffe                     & caffe          & C++              & 2020-02-13                 & 0       & 42856   & 0        & 42856     \\ \midrule
https://github.com/vgvassilev/cling               & cling          & C++              & 2023-07-18                 & 57      & 28342   & 0        & 28399     \\ \midrule
https://github.com/ipkn/crow                      & crow           & C++              & 2022-09-20                 & 0       & 1434    & 0        & 1434      \\ \midrule
https://github.com/fakeNetflix/facebook-repo-ds2  & ds2            & C++              & 2019-07-17                 & 367     & 27011   & 0        & 27378     \\ \midrule
https://github.com/freeminer/freeminer            & freeminer      & C++              & 2023-04-22                 & 15090   & 130828  & 1077     & 146995    \\ \midrule
https://github.com/meta-toolkit/meta              & meta           & C++              & 2017-08-19                 & 132     & 25451   & 0        & 25583     \\ \midrule
https://github.com/panda3d/panda3d                & panda3d        & C++              & 2023-06-13                 & 44671   & 416212  & 175      & 461058    \\ \midrule
https://github.com/facebook/proxygen              & proxygen       & C++              & 2023-07-20                 & 2676    & 70161   & 0        & 72837     \\ \midrule
https://github.com/s3fs-fuse/s3fs-fuse            & s3fs-fuse      & C++              & 2023-07-19                 & 197     & 19582   & 0        & 19779     \\ \midrule
https://github.com/cglib/cglib                    & cglib          & Java             & 2022-02-08                 & 0       & 0       & 15187    & 15187     \\ \midrule
https://github.com/deeplearning4j/deeplearning4j  & deeplearning4j & Java             & 2023-06-21                 & 0       & 224997  & 696717   & 921714    \\ \midrule
https://github.com/apache/drill                   & drill          & Java             & 2023-06-21                 & 538     & 34591   & 626295   & 661424    \\ \midrule
https://github.com/google/guava                   & guava          & Java             & 2023-07-18                 & 0       & 0       & 356651   & 356651    \\ \midrule
https://github.com/immutables/immutables          & immutables     & Java             & 2023-06-16                 & 0       & 0       & 69505    & 69505     \\ \midrule
https://github.com/dropwizard/metrics             & metrics        & Java             & 2023-07-20                 & 0       & 0       & 31317    & 31317     \\ \midrule
https://github.com/igniterealtime/Openfire        & Openfire       & Java             & 2023-07-20                 & 120     & 0       & 122186   & 122306    \\ \midrule
https://github.com/HubSpot/Singularity            & Singularity    & Java             & 2022-11-18                 & 0       & 0       & 122183   & 122183    \\ \midrule
https://github.com/igniterealtime/Smack           & Smack          & Java             & 2023-04-26                 & 0       & 0       & 125547   & 125547    \\ \midrule
https://github.com/igniterealtime/Spark           & Spark          & Java             & 2023-05-11                 & 9       & 0       & 91886    & 91895     \\ \midrule
TOTAL                                             &                &                  &                            & 6317493 & 1484547 & 2258726  & 10060766  \\ \bottomrule
\end{tabular}
}
\end{table}

\section{Methodology}\label{methodology}
For our study, identifiers are collected from and analyzed in the following contexts: class names, function names, parameter names, attribute names (i.e., data members), and declaration-statement names. A declaration-statement name is a name belonging to a local (to a function) or global variable. We use this terminology because it is consistent with srcML's terminology \cite{collard:2016} for these variables, and we used srcML to collect identifiers. Therefore, to study closed-category grammar patterns, we group identifiers based on these five categories. The purpose of this categorization is to examine the closed-category grammar patterns based on their high-level semantic roles (e.g., class names have a different role than function names). We collected these identifiers from 30 open-source systems, which are listed in Table \ref{tab:study_oss}. These systems belonged to a curated dataset of engineered software projects, synthesized by Reaper \cite{Munaiah2017}, a tool that measures how well different projects adhere to software engineering practices, such as documentation and continuous integration. 

The set of systems has an average and median of 335,358 and 111,069 LOC, respectively. 11 of the systems are primarily C systems, 9 are mainly C++, and 10 are primarily Java. We chose systems that have tests and use continuous integration (CI) under the idea that these represent systems with at least some basic process for ensuring quality; Reaper is able to automatically determine which systems have both CI and tests. Our primary concern in selecting systems is that they represent different programming languages, follow basic quality procedures, and are large enough for us to collect a sufficient number of identifiers. Given this, our choice of systems is designed to ensure that the grammar patterns in this study are applied across at least the languages under study.


\begin{table}[]
\centering
\caption{Distribution of part-of-speech labels in Old Data Set and CCID}
\label{tab:tag_distributions}
\begin{tabular}{@{}cccc@{}}
\toprule
\multicolumn{2}{c}{Old Data Set}                             & \multicolumn{2}{c}{CCID} \\ \midrule
\multicolumn{1}{l}{TAG} & \multicolumn{1}{l|}{FREQUENCY}     & TAG     & FREQUENCY      \\ \midrule
NM                      & \multicolumn{1}{c|}{1604 (45.2\%)} & N       & 1141 (31.58\%)  \\ \midrule
N                       & \multicolumn{1}{c|}{1141 (32.1\%)} & NM      & 643 (17.79\%)   \\ \midrule
V                       & \multicolumn{1}{c|}{305 (8.6\%)}   & P       & 398 (11.02\%)   \\ \midrule
NPL                     & \multicolumn{1}{c|}{238 (6.7\%)}   & V       & 363 (10.04\%)   \\ \midrule
PRE                     & \multicolumn{1}{c|}{105 (3\%)}     & DT      & 308 (8.52\%)    \\ \midrule
P                       & \multicolumn{1}{c|}{94 (2.6\%)}    & D       & 283 (7.83\%)    \\ \midrule
D                       & \multicolumn{1}{c|}{27 (0.8\%)}    & PRE     & 217 (6.00\%)      \\ \midrule
DT                      & \multicolumn{1}{c|}{15 (0.4\%)}    & NPL     & 142 (3.93\%)    \\ \midrule
VM                      & \multicolumn{1}{c|}{13 (0.4\%)}    & VM      & 69 (1.91\%)     \\ \midrule
CJ                      & \multicolumn{1}{c|}{8 (0.2\%)}     & CJ      & 50 (1.38\%)     \\ \midrule
Total                   & \multicolumn{1}{c|}{3550}          & Total   & 3614           \\ \bottomrule
\end{tabular}
\end{table}

\begin{table}[]
\centering
\caption{Distribution of Tags in Candidate and Verified (Manually-annotated) data set}
\label{tab:sample_distribution}
\resizebox{\columnwidth}{!}{%
\begin{tabular}{@{}lllll@{}}
\toprule
                                 & \multicolumn{2}{c}{\textbf{CJ}}                             & \multicolumn{2}{c}{\textbf{DT}}                   \\ \midrule
\multicolumn{1}{l|}{}            & Candidate               & \multicolumn{1}{l|}{Verified}     & Candidate               & Verified                \\ \midrule
\multicolumn{1}{l|}{Attribute}   & 66 (28.09\%)            & \multicolumn{1}{l|}{6 (12.24\%)}  & 78 (24.92\%)            & 84 (27.54\%)            \\
\multicolumn{1}{l|}{Declaration} & 62 (26.38\%)            & \multicolumn{1}{l|}{10 (20.41\%)} & 79 (25.24\%)            & 85 (27.87\%)            \\
\multicolumn{1}{l|}{Parameter}   & 44 (18.72\%)            & \multicolumn{1}{l|}{6 (12.24\%)}  & 78 (24.92\%)            & 58 (19.02\%)            \\
\multicolumn{1}{l|}{Function}    & 63 (26.81\%)            & \multicolumn{1}{l|}{27 (55.10\%)} & 78 (24.92\%)            & 78 (25.57\%)            \\
\multicolumn{1}{l|}{Class}       & 0 (0.00\%)              & \multicolumn{1}{l|}{0 (0.00\%)}   & 0 (0.00\%)              & 0 (0.00\%)              \\ \midrule
Total                            & \multicolumn{1}{c}{235} & \multicolumn{1}{c}{49}            & \multicolumn{1}{c}{313} & \multicolumn{1}{c}{305} \\ \midrule
                                 & \multicolumn{2}{c}{\textbf{D}}                              & \multicolumn{2}{c}{\textbf{P}}                    \\ \midrule
\multicolumn{1}{l|}{}            & Candidate               & \multicolumn{1}{l|}{Verified}     & Candidate               & Verified                \\ \midrule
\multicolumn{1}{l|}{Attribute}   & 80 (21.98\%)            & \multicolumn{1}{l|}{62 (23.40\%)} & 89 (24.45\%)            & 103 (26.96\%)           \\
\multicolumn{1}{l|}{Declaration} & 81 (22.25\%)            & \multicolumn{1}{l|}{70 (26.42\%)} & 88 (24.18\%)            & 73 (19.11\%)            \\
\multicolumn{1}{l|}{Parameter}   & 81 (22.25\%)            & \multicolumn{1}{l|}{77 (29.06\%)} & 89 (24.45\%)            & 60 (15.71\%)            \\
\multicolumn{1}{l|}{Function}    & 80 (21.98\%)            & \multicolumn{1}{l|}{41 (15.47\%)} & 88 (24.18\%)            & 140 (36.65\%)           \\
\multicolumn{1}{l|}{Class}       & 42 (11.54\%)            & \multicolumn{1}{l|}{15 (5.66\%)}  & 10 (2.75\%)             & 6 (1.57\%)              \\ \midrule
Total                            & \multicolumn{1}{c}{364} & \multicolumn{1}{c}{265}           & \multicolumn{1}{c}{364} & \multicolumn{1}{c}{382} \\ \bottomrule
\end{tabular}%
}
\end{table}

\begin{table}[]
\centering
\caption{Balanced population of identifiers per context}
\label{tab:context_population}
\begin{tabular}{@{}lc@{}}
\toprule
Context & \multicolumn{1}{l}{Sample Population} \\ \midrule
Attribute & 312 (24.47\%)\\
Declaration & 306 (24.00\%) \\
Function & 313 (24.55\%)\\
Class & 52 (4.08\%)\\
Parameter & 292 (22.90\%) \\ \midrule
Total & 1275 \\ \bottomrule
\end{tabular}%
\end{table}

\subsection{Detecting and Sampling Identifiers with Closed-Category Terms}\label{goldset_construction}

Sampling identifiers that contain closed-category terms is challenging for two reasons: (1) They are relatively uncommon in production code, and (2) many such terms are ambiguous without context, making automatic tagging difficult. Table \ref{tab:tag_distributions} shows the distribution of closed-category PoS tags present in a data set we constructed in prior work \cite{Newman_GP} versus the CCID, which was constructed explicitly to increase the population of closed-category PoS in the set. To address this, we implemented a two-phase sampling strategy: (1) filtering identifiers that \textit{potentially} contain closed-category terms into candidate sets, and (2) manually verifying and annotating a statistically representative sample.

\vspace{0.5em}
\noindent\textbf{Phase 1: Filtering Candidate Identifiers}

We began with the CCID corpus, which contains 279,000 unique identifiers from production code. To collect these 279K identifiers, we used the srcML identifier getter tool \footnote{https://github.com/SCANL/srcml\_identifier\_getter\_tool} on the srcML archives resulting from running srcML\cite{collard:2016} on the system repository directories (Table~\ref{tab:study_oss}). To identify candidate sets:

\begin{itemize}
    \item \textbf{Numerals (D):} We selected identifiers containing at least one digit, using Python's \texttt{isdigit()} functionality. Numerals are easier to detect automatically and are unambiguous in token form. However, there are cases where a numeral will be annotated as part of another category. For example, \textit{str2int} uses the numeral 2 as a preposition (to).
    
    \item \textbf{Determiners (DT), Conjunctions (CJ), and Prepositions (P):} We constructed lexicons for each category using curated lists of common English terms\footnote{\url{https://en.wikipedia.org/wiki/List_of_English_determiners}}\footnote{\url{https://www.englishclub.com/grammar/prepositions-list.php}}\footnote{\url{https://7esl.com/conjunctions-list/}}. We then filtered for identifiers containing component words (i.e., split tokens) that matched a word in one of these lists. This approach is viable only because these categories are closed and finite in vocabulary.
\end{itemize}
This filtering process produced the following candidate counts:
\begin{itemize}
    \item 602 identifiers candidate conjunctions
    \item 1,693 identifiers candidate determiners
    \item 3,383 identifiers candidate prepositions
    \item 4,630 identifiers candidate numerals
\end{itemize}

We use the term \textit{candidate} because these filters do not account for context or usage, and thus include false positives. Still, they serve as an upper bound for the prevalence of each category in the corpus. Based on this, we estimate the proportion of identifiers containing each term type as follows:
\begin{itemize}
    \item \textbf{0.2\%} (602/279,000) contain conjunctions
    \item \textbf{0.6\%} (1,693/279,000) contain determiners
    \item \textbf{1.2\%} (3,383/279,000) contain prepositions
    \item \textbf{1.7\%} (4,630/279,000) contain numerals
\end{itemize}

\vspace{0.5em}
\noindent\textbf{Phase 2: Balanced Sampling and Annotation}

Using a 95\% confidence level and a 5\% margin of error, we computed minimum sample sizes for each category. For example, a 95 and 5 sample for conjunctions (602 identifiers) is 235:
\begin{itemize}
    \item \textbf{CJ:} 235 candidate conjunction identifiers
    \item \textbf{DT:} 313 candidate determiner identifiers
    \item \textbf{P:} 345 candidate preposition identifiers
    \item \textbf{D:} 355 candidate numeral identifiers
\end{itemize}

Before manual annotation, we stratified the candidate identifiers by their program context:
\begin{itemize}
    \item Function names
    \item Parameters
    \item Attributes (i.e., class members)
    \item Function-local declarations
    \item Class names
\end{itemize}

In some contexts, such as parameters and especially class names, terms were underrepresented due to the natural scarcity of closed-category terms in those positions. To increase representation from underrepresented contexts, we attempted to oversample relevant subgroups. However, even with oversampling, the absolute number of qualifying identifiers (e.g., a parameter containing a conjunction) remained low. As our sampling was driven by the presence/population of closed-category terms in general, rather than population within our program contexts under study, we opted not to artificially balance the dataset further.

After sampling and stratification, we obtained a total of 1,275 identifiers across the four categories in our \textbf{candidate set}:
\begin{itemize}
    \item \textbf{364} candidate preposition identifiers
    \item \textbf{363} candidate numeral identifiers
    \item \textbf{313} candidate determiner identifiers
    \item \textbf{235} candidate conjunction identifiers
\end{itemize}

Following manual verification and annotation (described in Section \ref{manual_labeling}), we retained only those identifiers that were confirmed to contain closed-category terms. Table~\ref{tab:sample_distribution} summarizes both the sampled totals and the verified counts. The final dataset consists of 1,001 identifiers confirmed to include at least one closed-category term. \textbf{These identifiers comprise the CCID}. Table~\ref{tab:context_population} shows the CCID, but broken down by program context instead of closed-category type.

\vspace{0.5em}
\noindent\textbf{Annotation Notes}. Some tools exist for part-of-speech tagging of source code identifiers (e.g., the ensemble tagger from \cite{newman_ensemble}), but these are slow at scale and are trained on datasets that underrepresent closed-category terms. For example, in our prior dataset \cite{Newman_GP}, used to train the aforementioned tagging approach \cite{newman_ensemble}; conjunctions, determiners, and prepositions made up only 0.2\%, 0.4\%, and 2.6\% of tags, respectively. Thus, we found manual annotation necessary to ensure sufficient coverage and correctness for our study.

As we are primarily concerned with production code, and prior work shows that test name grammar patterns differ from production names \cite{peruma_test_name_gp}, we did not collect any identifiers containing the word `test', or that appeared in a clearly marked test file or directory. In addition, note that Table~\ref{tab:tag_distributions} counts tags at the \textit{word} level (e.g., CJ CJ N counts two CJ tags), whereas Table~\ref{tab:sample_distribution} counts tags at the \textit{identifier} level (e.g., one identifier with multiple CJ tags counts as one). This explains occasional mismatches between sampled and actual tag distributions.

\subsection{Manual Process for Annotating Part-of-Speech} \label{manual_labeling}
Initially, one author (annotator) is assigned annotate each identifier in the CCID with its grammar pattern. The annotator has experience annotating identifiers with part-of-speech (PoS) tags from prior work \cite{Newman_GP, peruma_test_name_gp}. The process is as follows: The annotator is given a split (using Spiral \cite{HuckaSpiral}) identifier, along with its type, file path, and line number, to facilitate easy identification of the identifier in the original code. The annotator is permitted to examine the source code from which the identifier originated, if necessary. The annotator is asked to additionally identify and correct mistakes made by Spiral. When the annotator is finished, two additional annotators are asked to validate (agree or disagree) with the annotations created by the original annotator. Any disagreements are discussed and fixed, if required. Furthermore, a fourth annotator assigned their own annotations, which are then compared to the original annotator's work. Again, disagreements are discussed and fixed.  An example disagreement is with the identifier \textit{where\_len}, which is a tricky one because `where' is typically an adverb or conjunction.  However, in this case `where' is a reference to a void pointer variable called `where' within the code. So `where\_len' is the length of the memory this pointer points to, making `where' a noun-adjunct in this case; describing the type of length. Thus, its grammar pattern is NM N. \textbf{The Fleiss' Kappa for this process was .916}.

We did not expand abbreviations for a couple reasons. The first is that some abbreviations are more meaningful than their expanded terms (e.g., HTTP, IPv4, SSL) due to their frequent use in abbreviated form by the community. The second reason is that abbreviation expansion techniques are not widely available and vary widely in terms of effectiveness on different types of terms \cite{newmanabbrev, Jingxuan:2023}. Therefore, a realistic worst-case scenario for developers and researchers is that no abbreviation-expansion technique is available to use, and their PoS taggers must work in this worst-case scenario. Whenever we recognize one, we do not split domain-term abbreviations (e.g., Spiral will make IPv4 into IPv 4; we corrected this to IPv4). We do this because it is the view of the authors that they should be recognized and appropriately tagged in their abbreviated (i.e., their most common) form.

\begin{table}[]
\centering
\caption{Most common Patterns}
\label{tab:MPC}
\resizebox{\textwidth}{!}{%
\begin{tabular}{@{}llllllll@{}}
\toprule
\multicolumn{2}{c}{\textbf{Determiner}} & \multicolumn{2}{c}{\textbf{Conjunction}} & \multicolumn{2}{c}{\textbf{Preposition}} & \multicolumn{2}{c}{\textbf{numeral}} \\ \midrule
DT N &
  \multicolumn{1}{l|}{109 (35.74\%)} &
  N CJ N &
  \multicolumn{1}{l|}{6 (12.24\%)} &
  P N &
  \multicolumn{1}{l|}{70 (18.32\%)} &
  N D &
  125 (47.17\%) \\ \midrule
DT NM N &
  \multicolumn{1}{l|}{40 (13.11\%)} &
  V CJ N &
  \multicolumn{1}{l|}{3 (6.12\%)} &
  P NM N &
  \multicolumn{1}{l|}{31 (8.12\%)} &
  NM N D &
  33 (12.45\%) \\ \midrule
DT NPL &
  \multicolumn{1}{l|}{20 (6.56\%)} &
  CJ NM &
  \multicolumn{1}{l|}{2 (4.08\%)} &
  P V &
  \multicolumn{1}{l|}{13 (3.40\%)} &
  N D N &
  9 (3.40\%) \\ \midrule
DT V &
  \multicolumn{1}{l|}{7 (2.30\%)} &
  NM CJ NM &
  \multicolumn{1}{l|}{2 (4.08\%)} &
  N P N &
  \multicolumn{1}{l|}{12 (3.14\%)} &
  PRE N D &
  8 (3.02\%) \\ \midrule
DT NM NM N &
  \multicolumn{1}{l|}{6 (1.97\%)} &
  V CJ V &
  \multicolumn{1}{l|}{2 (4.08\%)} &
  V P N &
  \multicolumn{1}{l|}{12 (3.14\%)} &
  V N D &
  8 (3.02\%) \\ \midrule
N DT &
  \multicolumn{1}{l|}{6 (1.97\%)} &
  V N CJ N &
  \multicolumn{1}{l|}{2 (4.08\%)} &
  P &
  \multicolumn{1}{l|}{10 (2.62\%)} &
  NPL D &
  6 (2.26\%) \\ \midrule
V DT &
  \multicolumn{1}{l|}{6 (1.97\%)} &
  CJ &
  \multicolumn{1}{l|}{1 (2.04\%)} &
  NM N P N &
  \multicolumn{1}{l|}{9 (2.36\%)} &
  N D NM N &
  3 (1.13\%) \\ \midrule
DT NM NPL &
  \multicolumn{1}{l|}{5 (1.64\%)} &
  CJ DT N &
  \multicolumn{1}{l|}{1 (2.04\%)} &
  V P &
  \multicolumn{1}{l|}{9 (2.36\%)} &
  V NM N D &
  3 (1.13\%) \\ \midrule
V DT N &
  \multicolumn{1}{l|}{5 (1.64\%)} &
  CJ NM N &
  \multicolumn{1}{l|}{1 (2.04\%)} &
  N P &
  \multicolumn{1}{l|}{8 (2.09\%)} &
  N D NPL &
  2 (0.75\%) \\ \midrule
DT &
  \multicolumn{1}{l|}{4 (1.31\%)} &
  CJ V &
  \multicolumn{1}{l|}{1 (2.04\%)} &
  NPL P N &
  \multicolumn{1}{l|}{8 (2.09\%)} &
  N P D NPL &
  2 (0.75\%) \\ \bottomrule
\end{tabular}%
}
\end{table}

\section{Evaluation of RQ1: \textbf{\RQA}} \label{rqadiscussion}
Our evaluation aims to establish, through RQ1 and RQ2: 1) how closed-category terms are used to convey differing types of program behavior, 2) the typical grammatical structure of identifiers containing closed-category terms, and 3) how closed-category term distributions differ across programming context, language, and system domains. This first research question investigates the semantic role of closed-category grammatical patterns in identifier naming. We focus on four closed-category part-of-speech types: prepositions, numerals, determiners, and conjunctions. We present our findings by (1) describing each category’s semantic function using axial codes, (2) summarizing behavioral trends via selective coding, and (3) highlighting shared trends through cross-category synthesis.

\subsection{Methodology: Manual Process for Behavioral Annotations}

We employed an approach inspired by Straussian grounded theory to analyze variable names in source code and their relationship to program behavior. This multi-phase coding process involved four annotators, combining individual annotations with iterative validation and synthesis to construct a theory grounded in observed naming patterns. The sample used in this study is a subset of the CCID described in Section \ref{methodology}. To construct this subset, we took the top 10 most common grammar patterns (Table \ref{tab:MPC}) and collected all identifiers that followed these patterns; randomly selecting the 10th grammar pattern if multiple patterns had the same frequency. These represent the most common names (i.e., from the perspective of grammatical structure) used in the data set. \textbf{This totals to 618 identifiers.}

Four annotators participated in the process, comprising both faculty and graduate students with prior experience in natural language processing and software engineering research. All annotators had previously worked on part-of-speech annotation tasks. Before formal annotation began, the team conducted a one-hour training and calibration session to discuss the guidelines, walk through examples, and establish expectations and deadlines.

\paragraph{\textbf{Coding Platform.}} Annotations are conducted collaboratively using a shared Google Sheets document. Each row in the sheet contained an identifier along with contextual metadata, including:

\begin{itemize}
    \item \textbf{Identifier Name}
    \item \textbf{Source Code Context}
    \item \textbf{Programming Language}
    \item \textbf{GitHub Commit Link}
    \item \textbf{Split Identifier Name} (tokenized form)
    \item \textbf{Grammar Pattern} (POS sequence)
    \item \textbf{Notes} (for open coding and memoing)
    \item \textbf{Axial Code} (for grouping behavioral patterns)
\end{itemize}

Open coding and memoing are captured directly in the \texttt{Notes} column. The final axial codes were recorded in the corresponding column once annotators had synthesized their observations.

\paragraph{\textbf{Phase 1: Familiarization.}}

All annotators reviewed the dataset to build familiarity with the variable names, associated grammar patterns, and program contexts. They discussed ambiguous or novel constructions in group chats to align interpretations and maintain consistency.

\paragraph{\textbf{Phase 2: Open Coding.}}

Annotators examined each variable name in its context and assigned a free-form behavioral interpretation based on how the variable is used in the surrounding code. These open codes and rationale are documented in the \texttt{Notes} column. The goal was to capture a grounded understanding of what each identifier conveyed, informed by both linguistic structure and program behavior.

\paragraph{\textbf{Phase 3: Axial Coding.}}

Annotators grouped similar open codes into higher-level axial codes, focusing on patterns where particular grammatical structures consistently aligned with specific behavioral roles. These axial codes captured mid-level abstractions (e.g., \textit{State Variables}, \textit{Event Triggers}), and were documented in the spreadsheet alongside notes justifying the grouping where needed. Each annotator’s axial codes were reviewed by a different annotator for validation. This cross-review process involved reading both the open codes and the proposed axial codes, discussing disagreements, and refining the categories until consensus was reached. The Fleiss' Kappa for this phase was: \textbf{.971 for numerals}, \textbf{.996 for Determiners}, \textbf{.976 for Prepositions}, and \textbf{1.0 for Conjunctions}.

\paragraph{\textbf{Phase 4: Selective Coding.}}

One annotator synthesized the final, validated axial codes across all annotations and constructed a set of selective codes representing core theoretical categories that linked grammar structure to program intent. These selective codes were then shared with the remaining annotators, who were asked to evaluate whether they reflected the themes and relationships they had observed during their own coding work. Annotators agreed or suggested revisions to finalize the theory.

\subsection{Numerals in Identifiers}

\paragraph{Overview.}
Numerals in identifiers act as compact, semantic indicators of structure, ordering, or version. They are also often used to disambiguate entities and encode numeric conventions. Their meaning is typically inferred through domain knowledge, making them powerful, but potentially hard to understand for those without the requisite domain knowledge.

\paragraph{Axial Codes.}
We created a dual-axis framework for interpreting the meaning of numerals, inspired by a single-axis framework we created in prior work on numerals in identifiers \cite{peruma_digits}. This framework reflects our observation that numerals contribute information in two distinct ways:
(1) the \textbf{role} they play within the local context (e.g., indexing, versioning), and
(2) the \textbf{source of meaning} they draw from, which is often external to the immediate source code scope (e.g., domain conventions, technical standards).
\textbf{Every numeral in the set has both a role and a source of meaning; they must be combined to fully understand the numeral.} We put an `x' between each combination of `Role' and `Source of Meaning' Axial Code.

\begin{itemize}
  \item \textbf{Role:} What functional purpose the numeral serves in the identifier.
  \begin{itemize}
    \item \textbf{Distinguisher:} The numeral differentiates conceptually similar entities, typically to avoid name collision errors from the compiler (e.g., \texttt{arg1}, \texttt{tile2}).
    \item \textbf{Version Identifier:} The numeral encodes versioning information such as protocol revisions or data format versions (e.g., \texttt{http2}, \texttt{v1}).
  \end{itemize}
  
  \item \textbf{Source of Meaning:} Where the interpretation of the numeral originates, typically via convention, tooling, or domain-specific logic.
  \begin{itemize}
    \item \textbf{Auto-Generated:} The numeral is added automatically by tools, compilers, or naming systems to avoid conflicts (e.g., \texttt{var1\_2}, \texttt{jButton3}).
    \item \textbf{Human-Named Convention:} The numeral's meaning is primarily derived from ad hoc developer intent and is not more complex than distinguishing entities manually (e.g., \texttt{str1}, \texttt{feature2}).
    \item \textbf{Locally Specific Concept:} The numeral conveys project- or context-specific information, often related to coordinate systems, data structures, or memory layouts (e.g., \texttt{m33} for matrix row 3 col 3).
    \item \textbf{Technology Term / Standard:} The numeral is part of a recognized domain-specific label, format, or protocol (e.g., \texttt{HTTP2}, \texttt{Neo4j}).
  \end{itemize}
\end{itemize}

\paragraph{Role x Source of Meaning}

\begin{enumerate}
\item \textbf{Distinguisher × Human-Named Convention (122 items)}\\
\textit{Description:} This group captures identifiers that use manually assigned numeric suffixes to distinguish conceptually and lexically similar entities.\\
\textit{Examples:} \texttt{host1} (first of 2 host variables), \texttt{e8} (element 8 in a parameter list)\\
\textit{Grammar patterns:}
\begin{itemize}
  \item N D (73)
  \item NM N D (21)
  \item V N D (6)
  \item NPL D (6)
  \item N D N (5)
  \item PRE N D (3)
  \item N D NM N (2)
  \item P D (2)
  \item PRE NM N D (2)
  \item V NM N D (2)
\end{itemize}

\item \textbf{Distinguisher × Locally Specific Concept (45 items)}\\
\textit{Description:} This group captures identifiers where numerals encode positional or logical roles based on system-specific conventions, such as grid layout or data structure indexing.\\
\textit{Examples:} \texttt{dist2} (squared distance calculation), \texttt{col1} (first column of a matrix)\\
\textit{Grammar patterns:}
\begin{itemize}
  \item N D (32)
  \item NM N D (5)
  \item PRE N D (3)
  \item N D N (2)
  \item NM N D P D (2)
  \item V N D (1)
\end{itemize}

\item \textbf{Distinguisher × Technology Term / Standard (17 items)}\\
\textit{Description:} This group captures identifiers that include numerals as part of standardized or domain-specific naming conventions, often encoding formats or specifications.\\
\textit{Examples:} \texttt{b1110} (binary for UTF8 byte sequences), \texttt{count32} (32-bit count value)\\
\textit{Grammar patterns:}
\begin{itemize}
  \item N D (7)
  \item NM N D (5)
  \item PRE N D (2)
  \item N D N (1)
  \item V N D (1)
  \item V NM N D (1)
\end{itemize}

\item \textbf{Version Identifier × Technology Term / Standard (9 items)}\\
\textit{Description:} This group captures identifiers where the numeral signals the version number of a protocol, tool, or technology component.\\
\textit{Examples:} \texttt{gw6} (gateway addr for IPV6), \texttt{httperf2} (version 2 of the \texttt{httperf} tool)\\
\textit{Grammar patterns:}
\begin{itemize}
  \item N D (5)
  \item NM N D (2)
  \item N D N (1)
  \item N D NM N (1)
\end{itemize}

\item \textbf{Distinguisher × Auto-Generated (8 items)}\\
\textit{Description:} This group captures identifiers that are automatically suffixed with a numeral to ensure uniqueness, often generated by tools or compilers.\\
\textit{Examples:} \texttt{field37}, \texttt{field4} (numbers are generated to avoid name collissions)\\
\textit{Grammar patterns:}
\begin{itemize}
  \item N D (8)
\end{itemize}
\end{enumerate}

\paragraph{Example.}
Consider the identifier \texttt{m34}, which appears in the context of a matrix operation. To fully interpret the numeral \texttt{3} in this name, we must consider both its Role and its Source of Meaning. Semantically, the numeral serves as a \textit{Distinguisher}; uniquely identifying this variable apart from its siblings (such as m32 and m31). However, its complete interpretation depends on its \textit{Locally Specific Concept} source: the developers have an internal convention that \texttt{3} refers to the row index, while \texttt{4} refers to the column. Without knowing the Source of Meaning, the numbers can only be interpreted as distinguishing one identifier from another; the meaning of the numerals would remain ambiguous. This illustrates how both axes work together—Role tells us \textit{what the numeral is doing}, while Source of Meaning tells us \textit{how to interpret the value}.

\paragraph{Selective Coding Insight.}
Numerals serve as semantic compression tools in source code: conveying versioning, layout, ordering, or configuration state using a minimal footprint. Their power lies in the idea that both the identifier's author and readers share a certain level of domain knowledge, and thus can understand the meaning of the numeral. Whether distinguishing hosts (\texttt{host1}, \texttt{host2}), signaling protocol versions (\texttt{http2}), or denoting matrix dimensions (\texttt{m33}), numerals rely on prior knowledge to be effective, this makes them:
\begin{itemize}
\item \textbf{Easy to understand} when used in well-known conventions (e.g., \texttt{3D}, \texttt{utf8})
\item \textbf{Hard to understand} when overused without documentation or when the reader lacks background information/experience that the author assumed they would have
\end{itemize}

Numerals are \textit{structural shortcuts} in the mental models of developers; a quick way to convey a lot of information in a small number of characters.

\subsection{Prepositions in Identifiers}
\paragraph{Overview.}
Prepositions in identifiers express spatial, temporal, or logical relationships. They are the most versatile (i.e., most axial codes) and frequently used closed-class grammatical structure in our dataset. Prepositions typically convey transformation, control conditions, event triggers, source origin, or context membership. Because only a subset of these are dual-axis (Boolean Flow), we inline the definitions with our examples, unlike with numerals, where we separate them. 

\paragraph{Axial Codes.}
Through axial coding, we identified several recurring behavioral roles that prepositions play in identifier names. These axial codes describe the functional semantics conveyed by the preposition within the naming context:

\begin{enumerate}
\item \textbf{Type Casting / Interpretation (38 items)}\\
\textit{Definition:} This group captures identifiers that signify transformation from one type, format, or abstraction to another.\\
\textit{Examples:} \texttt{str\_2\_int}, \texttt{as\_field}\\
\textit{Grammar patterns:}
\begin{itemize}
  \item P N (18)
  \item P NM N (9)
  \item N P N (4)
  \item V P N (2)
  \item P NM NM N (2)
  \item P V (1)
  \item NM N P N (1)
  \item V P (1)
\end{itemize}

\item \textbf{Position / Ordering in Time or Space (28 items)}\\
\textit{Definition:} This group captures identifiers that indicate relative position or sequencing within a spatial, temporal, or execution context.\\
\textit{Examples:} \texttt{before\_major}, \texttt{after\_first\_batch}\\
\textit{Grammar patterns:}
\begin{itemize}
  \item P N (8)
  \item P (4)
  \item N P N (3)
  \item P NM N (3)
  \item V P N (3)
  \item P V (2)
  \item V P (2)
  \item NM N P N (2)
  \item N P (1)
\end{itemize}

\item \textbf{Boolean Flow / Control Flag (26 items)}\\
\textit{Definition:} This group captures identifiers that encode boolean flags which both guard execution and describe the behavior they enable. This group is somewhat special, as their name implies other axial codes, but they are boolean variables. Thus, many of the identifiers in this group are dual-axis, where the 1st axis is boolean, and the 2nd is one of the other preposition axes. These variables are typically guards, used in branching logic that:
\begin{itemize}
  \item Activate based on \textit{position or sequencing} (e.g., \texttt{after\_equals})
  \item Govern \textit{strategy or type casting/interpretation behavior} (e.g., \texttt{for\_backprop}, \texttt{as\_array})
  \item Reflect \textit{data provenance or deferred logic} (e.g., \texttt{from\_docker\_config}, \texttt{wait\_for\_reload})
\end{itemize}
\textit{Examples:} \texttt{obsess\_over\_host}, \texttt{for\_backprop}\\
\textit{Grammar patterns:}
\begin{itemize}
  \item P N (12)
  \item P NM N (6)
  \item V P N (2)
  \item V P (2)
  \item N P N (2)
  \item N P (1)
  \item P (1)
\end{itemize}

\item \textbf{Data Source / Origin (20 items)}\\
\textit{Definition:} This group captures identifiers that refer to the source from which data or configuration is retrieved.\\
\textit{Examples:} \texttt{from\_context}, \texttt{from\_id}\\
\textit{Grammar patterns:}
\begin{itemize}
  \item P N (10)
  \item P NM N (1)
  \item N P N (2)
  \item P (3)
  \item NM N P N (1)
  \item V P (1)
  \item N P (2)
\end{itemize}

\item \textbf{Event Callback / Trigger (17 items)}\\
\textit{Definition:} This group captures identifiers that define behavior executed in response to user or system events.\\
\textit{Examples:} \texttt{on\_reason}, \texttt{on\_start}\\
\textit{Grammar patterns:}
\begin{itemize}
  \item P N (6)
  \item P NM N (5)
  \item P NM NM N (4)
  \item V P N (1)
  \item NM N P N (1)
\end{itemize}

\item \textbf{Deferred Processing / Pending Action (13 items)}\\
\textit{Definition:} This group captures identifiers that signal actions or data awaiting future handling.\\
\textit{Examples:} \texttt{to\_ack}, \texttt{to\_count}\\
\textit{Grammar patterns:}
\begin{itemize}
  \item P V (10)
  \item P N (2)
  \item P NM N (1)
\end{itemize}

\item \textbf{Unit-Based Decomposition / Measurement (11 items)}\\
\textit{Definition:} This group captures identifiers that describe per-unit measurement, processing, or aggregation.\\
\textit{Examples:} \texttt{down\_time}, \texttt{size\_in\_datum}\\
\textit{Grammar patterns:}
\begin{itemize}
  \item NPL P N (8)
  \item P N (1)
  \item N P N (1)
  \item NM N P N (1)
\end{itemize}

\item \textbf{Purpose / Role Annotation (10 items)}\\
\textit{Definition:} This group captures identifiers that clarify the functional role or use-case of a value.\\
\textit{Examples:} \texttt{for\_avg}, \texttt{for\_class}\\
\textit{Grammar patterns:}
\begin{itemize}
  \item P N (6)
  \item NM N P N (2)
  \item P NM N (1)
  \item V P (1)
\end{itemize}

\item \textbf{Data Movement / Transfer (9 items)}\\
\textit{Definition:} This group captures identifiers that represent movement of data or control between locations, buffers, or components.\\
\textit{Examples:} \texttt{to\_repo}, \texttt{to\_header}\\
\textit{Grammar patterns:}
\begin{itemize}
  \item P N (3)
  \item N P (3)
  \item P NM N (1)
  \item NM N P N (1)
  \item P NM NM N (1)
\end{itemize}

\item \textbf{Operation Basis / Strategy (8 items)}\\
\textit{Definition:} This group captures identifiers that describe the method, or trait that determines how operations may/should be carried out.\\
\textit{Examples:} \texttt{extend\_by\_hexahedron}, \texttt{with\_unary\_operator}\\
\textit{Grammar patterns:}
\begin{itemize}
  \item P N (2)
  \item P NM N (2)
  \item V P N (2)
  \item P (1)
  \item V P (1)
\end{itemize}

\item \textbf{Membership / Peer Grouping (7 items)}\\
\textit{Definition:} This group captures identifiers that signal inclusion in a group, scope, or set of peer entities.\\
\textit{Examples:} \texttt{in\_neighbour\_heap}, \texttt{in\_for}\\
\textit{Grammar patterns:}
\begin{itemize}
  \item P (2)
  \item P N (1)
  \item P NM N (1)
  \item V P N (1)
  \item V P (1)
  \item N P (1)
\end{itemize}

\item \textbf{Mathematical / Constraint Context (2 items)}\\
\textit{Definition:} This group captures identifiers that encode numerical limits, bounds, or ratios that constrain behavior.\\
\textit{Examples:} \texttt{over\_size}, \texttt{vmax\_over\_base}\\
\textit{Grammar patterns:}
\begin{itemize}
  \item P N (1)
  \item N P N (1)
\end{itemize}
\end{enumerate}

\paragraph{Selective Coding Insight.}
Prepositions in identifier names serve as compact, highly expressive relational markers. Across the dataset, prepositions consistently support four core semantic roles:

\begin{itemize}
\item \textbf{Transformation and Directionality}: Prepositions like \texttt{to}, \texttt{from}, and \texttt{as} signal type casting, movement, or format conversion.
\item \textbf{Execution and Conditional Control}: Prepositions such as \texttt{after}, \texttt{on}, and \texttt{for} often signal when or whether an action should occur, especially within event-driven operations and boolean flags that gate execution.
\item \textbf{Role and Configuration Semantics}: Prepositions like \texttt{with}, \texttt{by}, and \texttt{in} clarify how values contribute to a process or how behavior is scoped or grouped.
\item \textbf{Quantification and Unit-Based Aggregation}: Prepositions such as \texttt{per} and \texttt{in} describe how quantities are measured, normalized, or decomposed across units (e.g., \texttt{iterations\_per\_sample}, \texttt{size\_in\_datum}).
\item \textbf{Future-Intent or Deferred Action}: Especially with \texttt{to}, some identifiers encode pending or scheduled behavior (e.g., \texttt{to\_merge}, \texttt{wait\_for\_reload}).
\end{itemize}

Importantly, boolean flags that include prepositions do not form a distinct behavioral class, but instead overlay these four functions; gating type conversions, controlling source-based logic, or scoping strategies. These flags act as behavioral summaries, where the identifier directly reflects the guarded behavior (e.g., \texttt{send\_to\_buffer} reflects that the guarded code sends data to a buffer).

In short, prepositions make invisible system relationships visible. They map the logic of control, transformation, and association directly into identifier structure, enabling expressive, intention-revealing naming in complex systems.

\subsection{Determiners in Identifiers}

\paragraph{Overview.}
Determiners in identifiers help interpret values in relation to a set. They often signal positional reasoning, filtering criteria, relative thresholds, control flow, or scoping rules. In our analysis, we treat terms like \texttt{next} and \texttt{last} as \textbf{determiners}, even though they are typically categorized as adjectives in general English. In source code, however, these terms function more like determiners because \textbf{they specify a particular entity within a sequence or collection rather than merely describing its propertie}s. For example, the \texttt{next} pointer in a linked list does not describe a type of pointer, but rather identifies the specific node that follows in the structure. In this way, such terms serve a determinative function.

\paragraph{Axial Codes.}
We identified the following eight categories of determiner-based behavior:

\begin{enumerate}
\item \textbf{Temporal / Most Recent Element (60 items)}\\
\textit{Definition:} This group captures identifiers that refer to the most recently computed, stored, or observed value, often used for computing prior state, and in sequence-based data structures.\\
\textit{Examples:} \texttt{last\_bucket}, \texttt{last\_builder}\\
\textit{Grammar patterns:}
\begin{itemize}
  \item DT N (32)
  \item DT NM N (19)
  \item DT NM NM N (4)
  \item DT V (2)
  \item V DT N (2)
  \item DT NPL (1)
\end{itemize}

\item \textbf{Temporal / Upcoming Element (54 items)}\\
\textit{Definition:} This group captures identifiers that denote the next item in a sequence or timeline, often used in look-ahead and sequence-based data structures.\\
\textit{Examples:} \texttt{next\_tex}, \texttt{next\_bar}\\
\textit{Grammar patterns:}
\begin{itemize}
  \item DT N (35)
  \item DT NM N (9)
  \item DT V (3)
  \item N DT (3)
  \item DT NPL (2)
  \item DT NM NM N (1)
  \item V DT N (1)
\end{itemize}

\item \textbf{Population / Subpopulation Reference (42 items)}\\
\textit{Definition:} This group encompasses identifiers that reference a population or subset, typically using quantifiers such as \texttt{all}, \texttt{any}, or \texttt{some} to guide iteration, filtering, or policy logic.\\
\textit{Examples:} \texttt{any\_diffuse}, \texttt{all\_set}\\
\textit{Grammar patterns:}
\begin{itemize}
  \item DT NPL (13)
  \item DT N (9)
  \item V DT (6)
  \item DT NM NPL (4)
  \item V DT NPL (4)
  \item DT NM N (2)
  \item N DT (2)
  \item DT V (1)
  \item V DT N (1)
\end{itemize}

\item \textbf{Immediate Context Reference (26 items)}\\
\textit{Definition:} This group captures identifiers that refer to the current instance, scope, or runtime context—emphasizing locality, such as \texttt{this}, \texttt{another}, or \texttt{a}.\\
\textit{Examples:} \texttt{this\_node}, \texttt{another\_id}\\
\textit{Grammar patterns:}
\begin{itemize}
  \item DT N (17)
  \item DT NM N (6)
  \item DT NM NM N (1)
  \item N DT (1)
  \item V DT N (1)
\end{itemize}

\item \textbf{Negation / Exclusion Flag (18 items)}\\
\textit{Definition:} This group captures identifiers that indicate something is explicitly disabled, excluded, or absent; commonly using \texttt{no} to toggle features or signal null conditions.\\
\textit{Examples:} \texttt{no\_callback}, \texttt{no\_log}\\
\textit{Grammar patterns:}
\begin{itemize}
  \item DT N (12)
  \item DT NM N (2)
  \item DT NPL (2)
  \item DT NM NPL (1)
  \item DT V (1)
\end{itemize}

\item \textbf{Quantity Threshold / Optional Extensibility (4 items)}\\
\textit{Definition:} This group captures identifiers that express minimum thresholds, or the possibility of extending beyond a baseline.\\
\textit{Examples:} \texttt{enough\_memory}, \texttt{more\_data}\\
\textit{Grammar patterns:}
\begin{itemize}
  \item DT N (2)
  \item DT NPL (2)
\end{itemize}

\item \textbf{Default / Fallback Value Representation (2 items)}\\
\textit{Definition:} This group captures identifiers that represent placeholder or fallback values, used when a field must be filled or a default condition must be satisfied.\\
\textit{Examples:} \texttt{a\_void}, \texttt{no\_val}\\
\textit{Grammar patterns:}
\begin{itemize}
  \item DT N (2)
\end{itemize}

\item \textbf{Boolean Multi-Condition Test (2 items)}\\
\textit{Definition:} This group captures boolean identifiers representing conjunctions of multiple conditions, usually requiring all to be satisfied (e.g., both X and Y must be true).\\
\textit{Examples:} \texttt{both\_empty\_selection}, \texttt{both\_NonEmpty\_Selection}\\
\textit{Grammar patterns:}
\begin{itemize}
  \item DT NM N (2)
\end{itemize}
\end{enumerate}

\paragraph{Selective Coding Insight.}
Determiner-based identifiers help interpret values in relation to a set—by signaling \textit{position}, \textit{filtering criteria}, \textit{thresholds}, or \textit{scoping rules}. These are closed-category terms that enable programmers to express \textit{set logic}, \textit{entity selection}, and \textit{relative capacity or validity}. They typically support:
\begin{itemize}
\item \textbf{Positional reasoning} (\texttt{next}, \texttt{last}, \texttt{this}): Indicates where a value occurs in a temporal or structural sequence, helping to track state progression, history, or future execution.

\item \textbf{Population membership and filtering} (\texttt{some}, \texttt{any}, \texttt{each}, \texttt{least}, \texttt{which}): Refers to selecting or referencing members within a larger set, expressing scope, quantification, or comparison.

\item \textbf{Thresholding and extensibility} (\texttt{enough}, \texttt{more}, \texttt{additional}): Indicates whether a minimum condition is met or whether more values can be included beyond a base requirement.

\item \textbf{Identity negation or fallback} (\texttt{no}, \texttt{none}, \texttt{a}, \texttt{without}): Flags exclusion, absence, or placeholder values—often tied to feature toggles or default logic.

\end{itemize}

\subsection{Conjunctions in Identifiers}

\paragraph{Overview.}
Conjunction-based identifiers are rare but expressive. They signal compound behavior, dual-mode interfaces, or gated logic—often making hidden control flow or semantic relationships visible. Their rarity likely stems from the fact that developers often express conjunctions in logic rather than names. But when used, they highlight either an intent to emphasize control-flow behavior or to capture structural duality within a single name.

\paragraph{Axial Codes.}
We identified seven categories of conjunctional behavior, each reflecting a different type of pairing, conditionality, or combination:

\begin{enumerate}
\item \textbf{Data Pair / Composite Value (7 items)}\\
\textit{Definition:} This group captures identifiers that hold or refer to two values used together or in alternation, typically for a shared behavioral role or composite purpose.\\
\textit{Examples:} \texttt{data\_or\_diff}, \texttt{function\_and\_data}\\
\textit{Grammar patterns:}
\begin{itemize}
  \item N CJ N (6)
  \item V CJ N (1)
\end{itemize}

\item \textbf{Guarded Action / Conditional Enablement (6 items)}\\
\textit{Definition:} This group captures identifiers that encode actions gated by internal logic; executing only if a condition is satisfied. The conjunction expresses conditional enablement or guarded behavior.\\
\textit{Examples:} \texttt{if\_present}, \texttt{if\_unique}\\
\textit{Grammar patterns:}
\begin{itemize}
  \item CJ NM (2)
  \item V CJ N (2)
  \item V CJ V (1)
  \item V CJ VM P (1)
\end{itemize}

\item \textbf{Combined Action / Sequential Behavior (3 items)}\\
\textit{Definition:} This group captures identifiers that describe a sequence of operations performed together, often representing merged behaviors.\\
\textit{Examples:} \texttt{hash\_and\_save}, \texttt{print\_and\_free\_json}\\
\textit{Grammar patterns:}
\begin{itemize}
  \item V CJ V (1)
  \item V CJ V N (1)
  \item V N CJ N (1)
\end{itemize}

\item \textbf{Shared Interface for Alternatives (1 item)}\\
\textit{Definition:} This group captures identifiers that define a shared interface or behavior over mutually exclusive alternatives, with the conjunction indicating a choice, not a combination.\\
\textit{Example:} \texttt{generate\_key\_or\_iv}\\
\textit{Grammar pattern:}
\begin{itemize}
  \item V N CJ N (1)
\end{itemize}

\item \textbf{Combined Configuration / UI Concept (1 item)}\\
\textit{Definition:} This group captures identifiers that refer to compound interface or configuration concepts, often blending multiple traits into a unified design or behavioral setting.\\
\textit{Example:} \texttt{look\_and\_feel}\\
\textit{Grammar pattern:}
\begin{itemize}
  \item NM CJ NM (1)
\end{itemize}

\item \textbf{Boolean Concept Name (1 item)}\\
\textit{Definition:} This group captures identifiers that encode a named logical or boolean relationship, usually by treating the conjunction itself as a symbolic concept.\\
\textit{Example:} \texttt{and}\\
\textit{Grammar pattern:}
\begin{itemize}
  \item CJ (1)
\end{itemize}

\item \textbf{Boolean Multi-Condition Test (1 item)}\\
\textit{Definition:} This group captures identifiers that evaluate multiple conditions simultaneously; typically for readiness or validation checks, returning true only if all constraints are met.\\
\textit{Example:} \texttt{null\_or\_empty}\\
\textit{Grammar pattern:}
\begin{itemize}
  \item NM CJ NM (1)
\end{itemize}
\end{enumerate}

\paragraph{Selective Coding Insight.}
Conjunction-based identifiers are especially useful when modeling:

\begin{itemize}
\item \textbf{Duality}: Representing more than one entity or mode simultaneously (e.g., \texttt{input\_and\_output}, \texttt{key\_or\_iv})
\item \textbf{Mutual Exclusion}: Encoding choices between alternatives—only one active at a time (e.g., \texttt{stream\_or\_cache})
\item \textbf{Preconditions}: Embedding logic into the name that would otherwise be hidden in branching statements (e.g., \texttt{load\_if\_needed}, \texttt{trigger\_if\_active})
\end{itemize}

Conjunctions are the rarest category in our data, and while it is difficult to draw firm conclusions about them, it is clear that `and', `or', and `if' are go-to conjunctions, particularly for Data Pairs and Guarded actions.

\subsection{Cross-Category Synthesis}

Across numerals, determiners, prepositions, and conjunctions, developers use closed-class grammatical structures to encode compact, behavior-rich semantics in identifiers. While each PoS category exhibits distinct tendencies, analysis of grammar patterns reveals broader functional themes and stylistic consistencies across categories.

\paragraph{\textbf{Boolean Semantics and Execution Control}.}
Our first cross-category behavior is the use of closed-class elements to encode boolean conditions, execution control, or logical gating:
\begin{enumerate}
    \item Determiners such as \textit{no}, \textit{some}, \textit{this}, and \textit{both} signal presence, exclusion, or multi-condition boolean evaluation.
    \item When used as booleans, Prepositions like \textit{as}, and \textit{with} tend to guard sections of code that implement the behavior described in the identifier name.
    \item Conjunctions surface explicitly in guarded or compound logic names (e.g., load\_if\_enabled, both\_ready) using patterns like V CJ N, NM CJ NM.
\end{enumerate}

Interestingly, booleans appear in all three of these contexts, but each is a different flavor; a way of expressing behavior that is unique to the closed-category terms used in the boolean identifier.





\paragraph{\textbf{Control Flow and Event Signaling Across Categories}.}
Closed-category terms across all four categories reflect a tendency to encode temporal, reactive, or preconditioned behavior:
\begin{enumerate}
    \item Prepositions like on, before, after, and by appear in structures such as P N and V P N, signaling timing, triggers, or basis for operation.
    \item Conjunctions explicitly model control conditions (if, and) or mutual exclusivity (or), often appearing in V CJ N or N CJ N structures.
    \item Determiners frequently encode sequence through next and last, realized in DT N and DT NM N patterns.
    \item Numerals imply procedural differentiation (method1, step2) or timeline indexing when appearing in coordinated identifiers (m31, m32).
\end{enumerate}

These names act as micro-control structures, embedding state transitions and flow logic directly into identifier names to help the reader understand when or how an identifier will/should be used.

\paragraph{\textbf{Multi-Dimensional Semantic Layering}.}
Grammar pattern analysis highlights how identifiers stack multiple behavioral dimensions:
\begin{enumerate}
    \item Prepositions convey direction, transformation, measurement, and order

    \item Determiners convey selection, quantity, and scope

    \item Numerals embed indexing, uniqueness, and domain roles

    \item Conjunctions encode logic composition and structural alternatives.
\end{enumerate}

These layered forms serve as semantic shortcuts to convey complex behavior with minimal words. They compress conditions, transformations, order, and relationships into concise forms that aim to assist program comprehension and understanding without excess verbiage.

Finally, the \textbf{grammar patterns} observed across our axial codings provide structural insight into how behavioral semantics are composed. When the closed-category term appears as the first token in a grammar pattern, such as in \texttt{DT NM N} or \texttt{P NM N}, it typically modifies or qualifies a single operand, forming a unary relation (e.g., temporal status or transformation of a noun phrase). In contrast, when the closed term is flanked by open-class terms, such as in \texttt{N P N} or \texttt{N CJ N}, the structure reflects a binary relation: two operands connected through a behavioral or logical relationship (e.g., data flow or choice). 

By combining our axial and selective codes with these syntactic patterns, we gain a fuller picture of identifier meaning: the open-category terms indicate *which* entities are involved, while the closed-category term signals *how* they are related or behave with respect to one another.

\subsection{Summary of RQ1}
Through qualitative analysis of closed-category terms in identifiers, we have uncovered and explored how these compact grammatical forms play a central role in expressing program behavior. Each part-of-speech category contributes distinct semantic functions, ranging from transformation and scoping to control flow and logical composition.

Together, they reveal how developers construct concise, behavior-rich identifiers that encode structure, timing, intent, and logic. Whether signaling preconditions (\texttt{load\_if\_enabled}), alternatives (\texttt{data\_or\_diff}), state (\texttt{last\_bucket}), or structural roles (\texttt{col1}), these terms form a functional lexicon that bridges source code, cognition, and context.

\begin{table}[]
\centering
\caption{Top 10 terms per closed-category part-of-speech tag}
\label{tab:word_pos_count}
\resizebox{\columnwidth}{!}{%
\begin{tabular}{@{}l|l|l|l@{}}
\toprule
\textbf{Determiner}         & \textbf{Conjunction}      & \textbf{Numeral}           & \textbf{Preposition}        \\ \midrule
last (79, 25.65\%)          & and (18, 36.00\%)         & 1 (77, 27.21\%)          & to (76, 19.10\%)            \\
next (69, 22.40\%)          & if (16, 32.00\%)          & 2 (71, 25.09\%)          & for (37, 9.30\%)            \\
all (52, 16.88\%)           & or (13, 26.00\%)          & 0 (20, 7.07\%)           & on (36, 9.05\%)             \\
no (31, 10.06\%)            & than (1, 2.00\%)          & 3 (17, 6.01\%)           & as (35, 8.79\%)             \\
this (21, 6.82\%)           & since (1, 2.00\%)         & 4 (15, 5.30\%)           & from (33, 8.29\%)           \\
each (7, 2.27\%)            & when (1, 2.00\%)          & 16 (13, 4.59\%)          & in (25, 6.28\%)             \\
the (6, 1.95\%)             & -                         & 8 (8, 2.83\%)            & after (21, 5.28\%)          \\
more (5, 1.62\%)            & -                         & 6 (6, 2.12\%)            & 2 (to) (17, 4.27\%)              \\
a (4, 1.30\%)               & -                         & 64 (4, 1.41\%)           & of (15, 3.77\%)             \\
some (4, 1.30\%)            & -                         & 32 (3, 1.06\%)           & with (14, 3.52\%)           \\ \bottomrule
\end{tabular}%
}
\end{table}

\section{Evaluation of RQ2: \textbf{\RQB}} \label{rqbdiscussion}
One interesting aspect of \textit{closed-category terms} is that they appear in different contexts within source code with varying frequency. This variation provides insight into how developers use these terms to express different types of meaning. For RQ2, we investigate how closed-category terms correlate with three types of context: (1) the local programming context in which a variable is declared (e.g., \textsc{Function}, \textsc{Attribute}), (2) the programming language of the source code in which the identifier was found, and (3) the broader system-level domain of the software in which it appears (e.g., domain-specific vs general-purpose projects). This 3-way perspective allows us to examine both how these terms are used \textit{within} individual source code structures, between programming languages, and how they reflect distinctions \textit{across} different kinds of systems.

We begin by analyzing the distribution of four closed categories: prepositions, determiners, conjunctions, and numerals, across five programming contexts and three programming languages. We discuss which categories are most frequent in which contexts/languages and consider how those patterns may reflect the communicative goals of the developer. We then extend this analysis to system-level domain context, comparing the normalized frequency of closed-category term usage between domain-specific and general-purpose systems. 


\subsection{Closed-Category Term Usage Across Programming Contexts, Programming Languages, and System Domains}

We now examine how differing contexts and closed-category grammar patterns relate to one another, and whether programming language further conditions their usage. We begin by analyzing cross-language correlations in the usage of closed-category terms, followed by an exploration of correlations in how these terms are used across different program contexts. We provide Table~\ref{tab:word_pos_count}, which shows frequencies and percentages for PoS and terms, to help the reader understand what types of terms are most prevalent. However, for this research question, we rely primarily on Tables~\ref{tab:chisquared_language}, \ref{tab:adjustedpearsons_language}, \ref{tab:chisquared_context}, and \ref{tab:adjustedpearsons_context}, which present the results of our Pearson Chi-square tests and standardized Pearson residuals. Using these, we highlight common patterns, terms, and the contexts or languages to which these patterns are correlated.

\subsubsection{\textbf{Language-Specific Differences in Closed-Category Term Usage}} 

Starting with an analysis of closed category terms and programming language, our \textbf{null hypothesis} is that there is no relationship between identifiers containing closed category terms and the programming language in which they appear. Our \textbf{alternative hypothesis} is that there is a relationship between identifiers that contain closed category terms and the programming language.

\vspace{0.5em}
\noindent\textbf{Methodology.} To perform our Chi-Square test, we use the CCID described in Section~\ref{goldset_construction}. We count how many times each closed-category PoS appears in C++, Java, or C code by analyzing all 1,001 identifiers that contain closed-category terms. For example, we might find that there were 20 numerals in our data set found in C++ code, and 5 numerals in Java code. Once we have these frequencies, we apply the Chi-Square test and Standardized Pearson residuals with Bonferroni correction to determine overall significance and per-part-of-speech significance, respectively.

\begin{table}[]
\centering
\caption{Results of Pearson’s Chi-Squared Test. df = 6, $\alpha$ = 0.05, critical value = 12.592, test statistic = 4.291}
\label{tab:chisquared_language}
\begin{tabular}{@{}lllll@{}}
\toprule
                      & C        & C++      & Java     & Chi-square per row \\ \midrule
D                     & 0.451887 & 0.275300 & 1.282415 & 2.009603           \\
DT                    & 0.457607 & 0.036988 & 0.678404 & 1.172999           \\
P                     & 0.056554 & 0.015121 & 0.116729 & 0.188405           \\
CJ                    & 0.621604 & 0.157629 & 0.140876 & 0.920108           \\
Chi-square per column & 1.587652 & 0.485038 & 2.218424 & 4.291115           \\ \bottomrule
\end{tabular}%
\end{table}

\begin{table}[]
\centering
\caption{Standardized Pearson Residuals Results. With Bonferroni Correction, a significant result is $\alpha$ = 0.05/12 = 0.0042, which translates to a ±2.87 critical value}
\label{tab:adjustedpearsons_language}
\begin{tabular}{@{}llll@{}}
\toprule
   & C         & C++       & Java      \\ \midrule
D  & 0.953270  & 0.740778  & -1.649081 \\
DT & -0.986465 & -0.279221 & 1.233406  \\
P  & -0.367728 & -0.189311 & 0.542514  \\
CJ & 0.983056  & -0.492859 & -0.480581 \\ \bottomrule
\end{tabular}%
\end{table}

\vspace{0.5em}
\noindent\textbf{Results.} The Chi-square test for programming language (Table~\ref{tab:chisquared_language}) did not produce a statistically significant result. Thus, we do not reject the null hypothesis: \textbf{there is no strong evidence that closed-category tag usage differs significantly by programming language}. However, exploratory analysis of the Standardized Pearson residuals in Table~\ref{tab:adjustedpearsons_language} offers insight into modest trends worth noting:

\begin{itemize}
    \item \textbf{Numerals (D)} are modestly underrepresented in \textbf{Java} (residual = --1.65), suggesting a mild tendency to avoid numeric suffixes in Java naming.
    \item \textbf{Determiners (DT)} are slightly overrepresented in \textbf{Java} (residual = 1.23), potentially reflecting more frequent use of quantifying or contextual modifiers.
\end{itemize}


    

\paragraph{\textbf{Summary.}} While we did not find significant statistical evidence linking closed-category tag usage to programming language, the residual analysis and qualitative trends suggest mild idiomatic differences, particularly around numeral usage and determiner phrasing. For example, in our Axial Code data from RQ1, \textbf{Population/Subpopulation Reference} identifiers were found in Java (21, 50\%) and C++ (18, 42\%) more than in C(3, 7\%). These patterns may reflect broader stylistic conventions or design idioms of each language, but should be interpreted cautiously given the statistical outcome.

\begin{table}[]
\centering
\caption{Results of Pearson’s Chi-Squared Test. df = 12, $\alpha = 0.05$, critical value = 21.026, test statistic = \textbf{88.893567}.}
\label{tab:chisquared_context}
\resizebox{\columnwidth}{!}{%
\begin{tabular}{@{}lllllll@{}}
\toprule
                         & ATTRIBUTE & CLASS     & DECLARATION & FUNCTION   & PARAMETER & Chi-square per row \\ \midrule
D                        & 0.44932   & 16.031139 & 0.776138    & 15.916173  & 10.634467 & 43.807237          \\
DT                       & 0.511266  & 6.398601  & 2.14863     & 0.959251   & 0.171805  & 10.189553          \\
P                        & 0.332388  & 0.506133  & 3.498335    & 8.724009   & 3.63817   & 16.699033          \\
CJ                       & 3.366551  & 1.027972  & 0.233783    & 12.071429  & 1.498009  & 18.197744          \\
\textbf{Chi-square per column} & 4.659525  & 23.963845 & 6.656886    & 37.670861  & 15.942451 & \textbf{88.893567} \\ \bottomrule
\end{tabular}%
}
\end{table}

\begin{table}[]
\centering
\caption{Standardized Pearson Residuals. With Bonferroni correction, a significant result is $\alpha = 0.05/20 = 0.0025$, which translates to a $\pm 3.02$ critical value.}
\label{tab:adjustedpearsons_context}
\begin{tabular}{@{}llllll@{}}
\toprule
    & ATTRIBUTE & CLASS             & DECLARATION & FUNCTION           & PARAMETER         \\ \midrule
D   & -0.905532 & \textbf{4.719156} & 1.1768      & \textbf{-5.505053} & \textbf{4.254103} \\
DT  & 0.993307  & \textbf{-3.065907} & 2.013483    & -1.389769          & -0.556035         \\
P   & 0.849262  & -0.91434          & -2.724314   & \textbf{4.444202}  & -2.713221         \\
CJ  & -2.179409 & -1.050735         & -0.567884   & \textbf{4.21543}   & -1.403873         \\ \bottomrule
\end{tabular}
\end{table}

\subsubsection{\textbf{Context-Specific Differences in Closed-Category Term Usage}}

Next, we analyze the correlation between closed-category terms and program contexts such as Function names, Attributes, and Class names. Our \textbf{null hypothesis} is that there is no relationship between identifiers containing closed-category terms and the context in which they appear. Our \textbf{alternative hypothesis} is that there is a relationship between identifiers containing closed-category terms and the context in which they appear.

\vspace{0.5em}
\noindent\textbf{Methodology.} To perform our Chi-Square test, we use the CCID described in Section~\ref{goldset_construction}. We analyzed all 1,001 identifiers that contained closed-category terms, and counted how many times a closed-category term appears in one of our five code contexts: Attribute, Function, Class, Declaration, or Parameter. Once we have these frequencies, we apply the Chi-Square test and Standardized Pearson residuals with Bonferroni correction to determine overall significance and per-part-of-speech significance, respectively.

\vspace{0.5em}
\noindent\textbf{Results.} The Chi-squared test for context (Table~\ref{tab:chisquared_context}) shows a significant result (88.89~$>$~21.026), allowing us to \textbf{reject the null hypothesis}. As before, we analyze the Standardized Pearson Residuals (Table~\ref{tab:adjustedpearsons_context}) to understand where the largest deviations appeared.

\textbf{Conjunctions (CJ)}. Closed-category grammar patterns that include conjunctions typically feature the terms `and', `or', or `if', as reflected in Table~\ref{tab:word_pos_count}. Although rare overall, these patterns are significantly positively correlated with function names (Standardized Pearson residual = 4.22, Table~\ref{tab:adjustedpearsons_context}). This indicates that when conjunctions do appear, they are far more likely to occur in function names than in other contexts.

The selective coding data from RQ1 explains this pattern. Conjunction-based grammar patterns tend to express compound logic, dual-purpose behavior, or guarded activation, which are most relevant when naming behaviors or actions rather than static values. For example, \textbf{Guarded Action / Conditional Enablement} patterns such as load\_if\_needed or activate\_if\_enabled appear in function names to encode preconditions or gating logic directly into the identifier.

Conjunction-based names are largely absent from declarations and classes, likely because those contexts do not typically represent conditional or compound operations. The data supports the interpretation that developers strategically use conjunctions in function names to foreground complex control logic or behavioral nuance at the point of execution.

While we did find a significant correlation between conjunctions and functions, it is essential to note that our dataset contains 50 conjunctions, meaning that while we have identified potential trends, further research on a larger sample is likely necessary.

\textbf{Determiners (DT)}. Closed-category grammar patterns that include determiners typically feature the terms \texttt{last}, \texttt{next}, \texttt{all}, \texttt{no}, or \texttt{this}, as shown in Table~\ref{tab:word_pos_count}. While determiners are not significantly positively correlated with any specific context, they are modestly negatively correlated with class names (Standardized Pearson residual = –3.07, Table~\ref{tab:adjustedpearsons_context}), suggesting that developers tend to avoid determiner-based grammar patterns in class names.

The selective coding analysis offers a plausible explanation: the most common roles for determiners involve expressing temporal or positional relationships—such as \textbf{Temporal / Most Recent Element} and \textbf{Temporal / Upcoming Element} (over 100 instances)—as well as set-based semantics, such as \textbf{Population / Subpopulation Reference} (38 instances). These patterns commonly use terms like \texttt{next}, \texttt{last}, \texttt{prev}, \texttt{all}, and \texttt{any} to indicate an element’s position in a sequence or its membership in a filtered subset.

These naming strategies are well-suited to attributes, parameters, and declarations, where variables often represent dynamic state or bounded subsets. In contrast, class names are generally used to describe abstract data types or roles, where positional or filtering semantics are less relevant. The relative absence of determiners in class contexts thus reflects their semantic focus: determiners foreground state, scope, or specificity, whereas class names typically signal structural purpose or generalization.

\textbf{Numerals (D)}. Closed-category grammar patterns that include numerals often feature numerals such as \texttt{1}, \texttt{2}, \texttt{0}, \texttt{3}, and \texttt{4}, as shown in Table~\ref{tab:word_pos_count}. Numerals are significantly positively correlated with parameter names and class names, and significantly negatively correlated with function names (Standardized Pearson residual = 4.72 for class names, 4.25 for parameters, and –5.51 for functions; Table~\ref{tab:adjustedpearsons_context}). Notably, numerals are the only closed-category category to exhibit a positive correlation with class names.

The selective coding data sheds light on this trend. The most frequent numeral-related patterns in our dataset fall under \textbf{Distinguisher × Human-Named Convention} and \textbf{Distinguisher × Locally Specific Concept}. These naming strategies are used to distinguish among similar entities (e.g., \texttt{arg1}, \texttt{arg2}, \texttt{tile3}) or to embed system-specific references (e.g., \texttt{m34}, \texttt{cp437}) into variable or type names. Such distinctions are especially useful in parameters and declarations, where there is no syntactic support for disambiguation outside of naming.

In contrast, function-level disambiguation is often handled by the language itself, through overloading, polymorphism, or naming conventions focused on behavior; making numerals largely unnecessary or even undesirable in that context. Their absence from function names reflects this shift: numerals encode identity, that is, they serve as a means of traceability to specific domain concepts and distinguish entities with similar names, rather than encoding purpose or behavior.

Taken together, these findings suggest that numerals serve primarily as disambiguators or protocol markers rather than communicative devices for expressing behavior. Their presence in class and parameter names signals static or structural variation, while their avoidance in function names underscores a developer's preference for meaningful, descriptive action labels over numerical markers.

\textbf{Prepositions (P)}. Closed-category grammar patterns that include prepositions frequently feature terms such as \texttt{to}, \texttt{for}, \texttt{as}, \texttt{on}, or \texttt{from}, as shown in Table~\ref{tab:word_pos_count}. These patterns are significantly positively correlated with function names (Standardized Pearson residual = 4.44, Table~\ref{tab:adjustedpearsons_context}), suggesting that developers are particularly likely to use prepositional grammar when naming behaviors or operations.

This strong correlation reflects the behavioral semantics that prepositions convey in identifier names. As detailed in our selective coding, prepositions frequently express directionality, transformation, conditional activation, or event-driven execution; all of which are highly function-oriented behaviors, requiring action to be taken. 


\paragraph{\textbf{Summary.}} The results of our analysis support the alternative hypothesis: closed-category parts of speech are meaningfully correlated with specific roles and contexts in source code. Prepositions and conjunctions appear more frequently in function names, where they help express behavioral nuances such as guarded actions, type casting, or alternative execution paths. Numerals, by contrast, are most commonly found in class names and parameter declarations, where they signal disambiguation, indexing, or versioning; identifiers rooted in identity rather than behavior. 


\begin{figure}[t]
  \centering
  \includegraphics[width=0.85\linewidth]{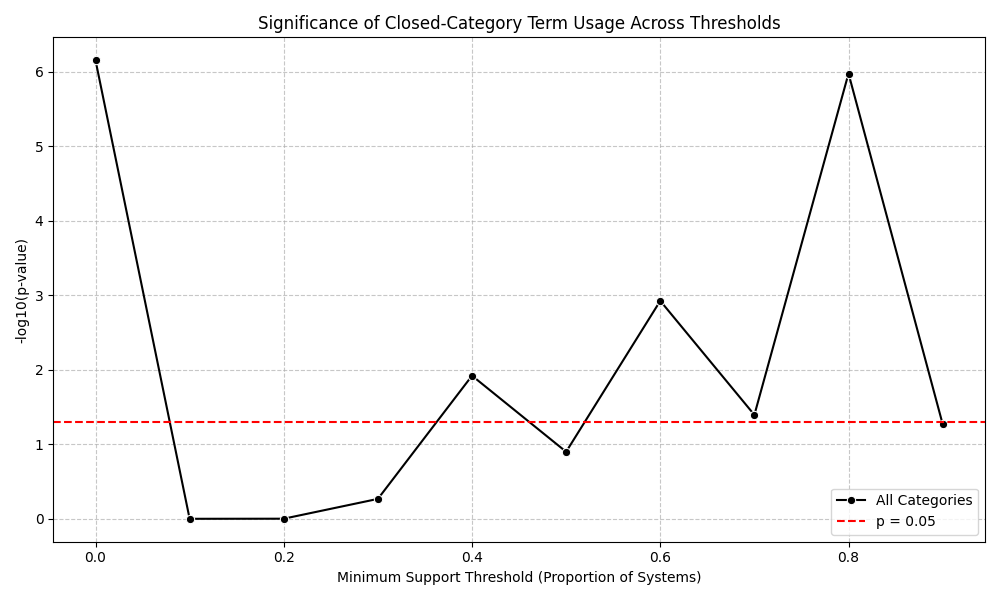}
  \caption{Global Mann-Whitney U test significance across thresholds, showing divergence between domain-specific and general systems. Peaks at 0.6 and 0.8 suggest the importance of both ubiquitous and moderately specific closed-category terms.}
  \label{fig:threshold_significance_global}
\end{figure}

\begin{figure}[t]
  \centering
  \includegraphics[width=0.85\linewidth]{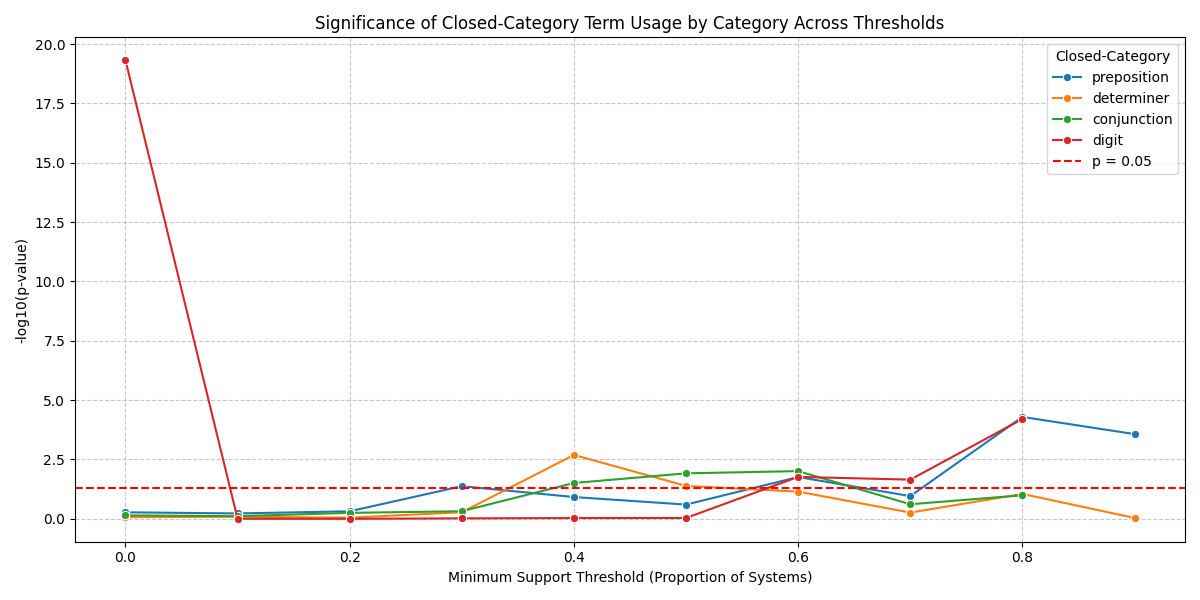}
  \caption{Per-category Mann-Whitney U test significance across thresholds. Prepositions dominate across thresholds, while conjunctions and numerals contribute more variably.}
  \label{fig:threshold_significance_per_category}
\end{figure}

\begin{figure}[t]
  \centering
  \includegraphics[width=0.85\linewidth]{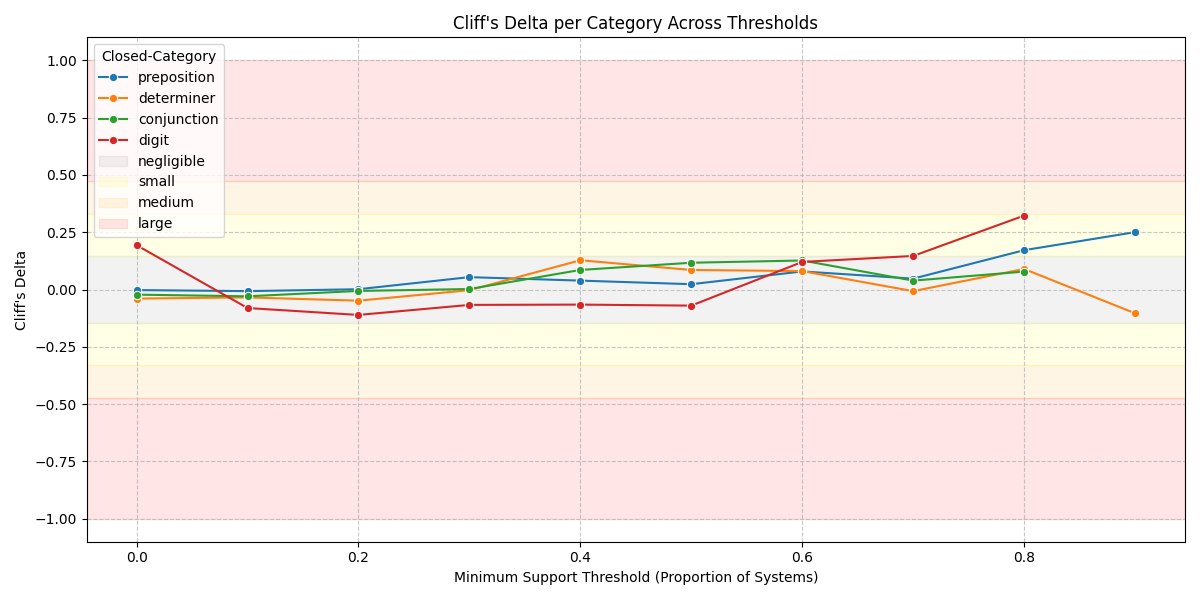}
  \caption{Cliff’s Delta for closed-category terms across system support thresholds.}
  \label{fig:cliffs_delta_per_threshold}
\end{figure}
\begin{table}[]
\centering
\caption{Systems and System Domains Selected Based on Axial Codes for each Closed-Category Type}
\label{tab:system_domains}
\resizebox{\columnwidth}{!}{%
\begin{tabular}{@{}lllll@{}}
\toprule
Closed-Category Type & Axial Code                                           & Domain                                         & GitHub URL                                         & Language \\ \midrule
Prepositions         & Type Casting / Interpretation                        & Serialization/Deserialization Libraries        & https://github.com/msgpack/msgpack-c/              & C        \\ \midrule
Prepositions         & Type Casting / Interpretation                        & Serialization/Deserialization Libraries        & https://github.com/open-source-parsers/jsoncpp.git & C++      \\ \midrule
Prepositions         & Type Casting / Interpretation                        & Polyglot Interop Tools / Type Bridge Layers    & https://github.com/pybind/pybind11                 & C++      \\ \midrule
Prepositions         & Type Casting / Interpretation                        & Polyglot Interop Tools / Type Bridge Layers    & https://github.com/swig/swig                       & C++      \\ \midrule
Determiners          & Position in Sequence (Temporal / Recent \& Upcoming) & Parser Generators / Token Stream Libraries     & https://github.com/akimd/bison                     & C        \\ \midrule
Determiners          & Position in Sequence (Temporal / Recent \& Upcoming) & Parser Generators / Token Stream Libraries     & https://github.com/antlr/antlr4                    & Java     \\ \midrule
Determiners          & Position in Sequence (Temporal / Recent \& Upcoming) & Job Queues / Schedulers                        & https://github.com/PerMalmberg/libcron             & C++      \\ \midrule
Determiners          & Position in Sequence (Temporal / Recent \& Upcoming) & Job Queues / Schedulers                        & https://github.com/quartz-scheduler/quartz         & Java     \\ \midrule
Prepositions         & Position / Ordering in Time or Space                 & Data Structure / Algorithm Libraries           & https://github.com/apache/commons-collections      & Java     \\ \midrule
Prepositions         & Position / Ordering in Time or Space                 & Data Structure / Algorithm Libraries           & https://github.com/boostorg/container              & C++      \\ \midrule
Prepositions         & Position / Ordering in Time or Space                 & Compiler / Intermediate Representation Tools   & https://github.com/rose-compiler/rose              & C        \\ \midrule
Prepositions         & Position / Ordering in Time or Space                 & Compiler / Intermediate Representation Tools   & https://github.com/TinyCC/tinycc                   & C        \\ \midrule
Determiners          & Population / Subpopulation Reference                 & Dataframe / Matrix Libraries                   & https://github.com/apache/arrow                    & C++      \\ \midrule
Determiners          & Population / Subpopulation Reference                 & ML Preprocessing / Feature Selection           & https://github.com/haifengl/smile                  & Java     \\ \midrule
Determiners          & Population / Subpopulation Reference                 & Dataframe / Matrix Libraries                   & https://github.com/jtablesaw/tablesaw              & Java     \\ \midrule
Determiners          & Population / Subpopulation Reference                 & ML Preprocessing / Feature Selection           & https://github.com/pjreddie/darknet                & C        \\ \midrule
Conjunctions         & Guarded Action / Conditional Enablement              & UI Libraries / Event Dispatch Systems          & https://github.com/GNOME/gtk                       & C        \\ \midrule
Conjunctions         & Guarded Action / Conditional Enablement              & Feature Flag Systems / Config-Driven Execution & https://github.com/lightbend/config                & Java     \\ \midrule
Conjunctions         & Guarded Action / Conditional Enablement              & UI Libraries / Event Dispatch Systems          & https://github.com/ocornut/imgui                   & C++      \\ \midrule
Conjunctions         & Guarded Action / Conditional Enablement              & Feature Flag Systems / Config-Driven Execution & https://github.com/spring-projects/spring-boot     & Java     \\ \midrule
Digits               & Distinguisher × Locally Specific Concept             & Game Engines / Grid-based Games                & https://github.com/godotengine/godot               & C++      \\ \midrule
Digits               & Distinguisher × Locally Specific Concept             & Game Engines / Grid-based Games                & https://github.com/jMonkeyEngine/jmonkeyengine     & Java     \\ \midrule
Digits               & Distinguisher × Locally Specific Concept             & Scientific Computing / Matrix Libraries        & https://github.com/OpenMathLib/OpenBLAS            & C        \\ \midrule
Digits               & Distinguisher × Locally Specific Concept             & Scientific Computing / Matrix Libraries        & https://github.com/PX4/eigen                       & C++      \\ \midrule
Digits               & Distinguisher × Human-Named Convention               & Code Generators / Macro Frameworks             & https://github.com/jhipster/jhipster-bom           & Java     \\ \midrule
Digits               & Distinguisher × Human-Named Convention               & GUI Builders / Form Designers                  & https://github.com/qt/qtbase                       & C++      \\ \midrule
Conjunctions         & Data Pair / Composite Value                          & Multi-format I/O Libraries                     & https://github.com/apache/parquet-java             & Java     \\ \midrule
Conjunctions         & Data Pair / Composite Value                          & Cryptographic Libraries                        & https://github.com/jedisct1/libsodium              & C        \\ \midrule
Conjunctions         & Data Pair / Composite Value                          & Multi-format I/O Libraries                     & https://github.com/libjpeg-turbo/libjpeg-turbo     & C        \\ \midrule
Conjunctions         & Data Pair / Composite Value                          & Cryptographic Libraries                        & https://github.com/openssl/openssl                 & C        \\ \bottomrule
\end{tabular}
}
\end{table}
\subsubsection{\textbf{Closed-Category Term Usage Across System Domains}}
Having established correlations between closed-category terms, source code context, and programming language, we now turn to a broader question: do these terms also vary with the domain of the software system itself? This sub-question allows us to further test our central hypothesis, that closed-category terms are not used arbitrarily, but instead reflect domain-relevant distinctions in how behavior and structure are communicated. If certain domains make more frequent or specialized use of closed-category terms, this suggests that such terms play a role in expressing concepts tightly coupled to those domains. Understanding and appropriately using these terms may therefore be critical for accurate communication of behavior in domain-specific software.

\vspace{0.5em}
\noindent\textbf{Methodology.} In RQ1, we developed a set of Axial Codes to describe the behavioral roles of closed-category terms in identifiers. To explore their importance at the level of system domain, we selected the two most common Axial Codes from each closed-category group (e.g., Prepositions, Determiners). For each code, we identified two software domains that we hypothesized would frequently use identifiers expressing that behavior. For example, in the Preposition group, the top two Axial Codes were:
\begin{itemize}
    \item \textbf{Type Casting / Interpretation}
    \item \textbf{Position / Ordering in Time / Space / Execution Context}
\end{itemize}

Based on these, we selected four relevant software domains:
\begin{itemize}
    \item For \textbf{Type Casting / Interpretation}: 
    \begin{itemize}
        \item Serialization/deserialization libraries
        \item Polyglot interop tools or type bridge layers
    \end{itemize}
    \item For \textbf{Position / Ordering in Time / Space / Execution Context}:
    \begin{itemize}
        \item Data structure and algorithm libraries
        \item Compiler or intermediate representation (IR) tooling
    \end{itemize}
\end{itemize}

Table~\ref{tab:system_domains} lists all selected systems, the domain they represent, and the Axial Code that motivated their inclusion. To fit the table, we omitted a few details like system size; this information can be found in our open data set (Section \ref{data_avail}). We analyzed identifiers drawn from five programming contexts—attributes, parameters, functions, declarations, and class names—across two groups of systems: one curated for domain-specific relevance and one composed of general-purpose projects, which were used to construct the CCID (Table~\ref{tab:study_oss}). The general-purpose group serves as a baseline, as these systems were not selected based on any particular domain hypothesis. Our underlying assumption is that, if closed-category terms are meaningfully correlated with domain-specific concerns, we will observe statistically significant differences in their usage between these two groups.

For each system, we extracted all identifiers and segmented them using Spiral~\cite{HuckaSpiral}. We then filtered out all terms that are neither numerals nor included in our predefined lists of closed-category terms (as defined in Section \ref{methodology}). After filtering, we compute the normalized frequency of closed-category term usage by dividing the count of qualifying terms by the system’s total lines of code. To assess whether the differences in usage were statistically significant, we applied a Mann-Whitney U test to compare the distributions between domain-specific and general-purpose systems.

\vspace{0.5em}
\noindent\textbf{Results.} To mitigate the risk of a small number of systems dominating the term distribution, and to better understand how widely closed-category terms are used, we introduce a support threshold that controls how many systems a term must appear in to be included in the Mann-Whitney U test. Increasing the threshold emphasizes more widely used (\textit{ubiquitous}) closed-category terms; decreasing it emphasizes more narrowly distributed (\textit{specific}) closed-category terms that may signal domain-specific behavior. Significance at high thresholds implies that there are terms that are important to all of our domain-specific systems; significance at lower thresholds implies that there are subsets of the domain-systems that make use of terms that are not very universal, but nevertheless set these systems apart from the general set.

To explore how these different usage profiles affect our results, we conducted a threshold sweep. At each level, a term had to appear in at least a given proportion of systems to be retained. This allowed us to systematically vary our emphasis between ubiquity (terms common across many systems) and specificity (terms concentrated in a smaller, domain-aligned subset). The results, shown in Figure~\ref{fig:threshold_significance_global}, reveal the most substantial distributional divergence at thresholds around 0.6 and 0.8. These peaks suggest that both common and moderately specific terms help distinguish domain-specific systems. By contrast, thresholds between 0.1 and 0.3 yielded little significance, likely reflecting linguistic noise from terms with low usage or ambiguous semantic function.

We repeat the analysis at the level of individual closed-category types (prepositions, determiners, conjunctions, numerals) to identify which groups drive the observed differences. As shown in Figure~\ref{fig:threshold_significance_per_category}, prepositions exhibit consistently strong significance across thresholds, particularly between 0.6 and 0.8. Numerals and conjunctions show more variable but still notable divergence, while determiners contribute the weakest and least consistent signal. These trends suggest that domain-specific systems rely more heavily on certain linguistic forms, especially prepositions, to express structural or behavioral distinctions central to their design.

To complement the significance testing, we examine \textit{Cliff's Delta} as a non-parametric effect size estimate, plotted in Figure~\ref{fig:cliffs_delta_per_threshold}. This allows us to assess not only whether closed-category usage differs between system types, but also how strongly. The results show that \textbf{prepositions} and \textbf{numerals} increasingly favor domain-specific systems at higher thresholds, reaching small to medium effect sizes. \textbf{Determiners}, by contrast, exhibit weak or even negative effect sizes, suggesting a more general-purpose usage profile. \textbf{Conjunctions} remain close to the negligible–small range, with mild domain skew. These patterns reinforce the idea that domain-specific systems do not just differ in which closed-category terms they use, but in how salient those terms are among their most widely reused identifiers.

\paragraph{\textbf{Summary.}} Our findings suggest that domain-specific systems tend to use closed-category terms more frequently than general-purpose baselines, particularly in ways that align with the communicative roles captured by our Selective Codes. While we rely on predefined lists of closed-category terms—without verifying each term's function in context—our goal in this evaluation was not to establish definitive usage, but to assess whether these terms might play a heightened role in domain-specific software. The statistical results support that possibility. As such, we argue that further research into how closed-category terms contribute to domain-specific expression is both warranted and promising. These findings offer initial evidence that supporting developers in the effective use of such terms could benefit certain styles or domains of software development.

\subsection{Summary of RQ2}

For RQ2, we examine how closed-category terms correlate with multiple forms of context: (1) source-code-local structure, (2) programming language, and (3) broader system domain. Our findings reveal several consistent trends. First, there is no statistically significant difference in the distribution of closed-category terms across the three programming languages under study, though there are some trends that indicate how they may differ in minor (i.e., non-statistically-significant) ways. Second, source code context plays a significant role: Prepositions and conjunctions are used disproportionately in function names; numerals are \textbf{significantly positively} correlated with parameters and class names while \textbf{significantly negatively} correlated with function names; and Determiners are significantly negatively correlated with class names. These patterns align with the communicative roles uncovered in our Selective Codes, such as the use of prepositions to express behavior or data flow, and numerals to distinguish instances or versions.

Finally, we found statistical evidence that domain-specific systems use closed-category terms more frequently than general-purpose ones. This suggests that these terms serve as meaningful signals of domain-relevant behavior. Taken together, our results demonstrate that \textbf{closed-category terms have specific, purposeful usage in software development}.

One of the broader aims of RQ2 is to assess whether closed-category terms are meaningful enough to warrant a dedicated study. We argue that their statistically significant correlations with specific code contexts support the need for further research: such terms appear deliberately and consistently in ways that reflect their natural language functions. While our domain-level comparison relies on predefined lists of closed-category terms, without manual verification of each term’s grammatical role, the results nonetheless suggest that these terms may hold particular communicative importance in domain-specific software. Supporting the appropriate use of closed-category terms through tools, naming conventions, or educational interventions may ultimately benefit program comprehension and internal quality, particularly in domains where such terms help convey behavioral intent.

\section{Related Work}\label{sec:related}
While numerous studies have been conducted on identifier names, this paper represents one of the few to address closed-category terms, and the only paper to conduct an in-depth analysis of their usage in open-source systems.  We discuss relevant related literature below, and how our work can be improved by, or improve upon, their outcomes.

\subsection{Grounded Theory in Software Engineering}
Our methodology is inspired by Straussian grounded theory \cite{CorbinStrauss:1990}, which emphasizes iterative coding, constant comparison, and the development of conceptual categories grounded in data. While we drew inspiration from classic Straussian grounded theory, our approach adapts it to the structure of source code, treating identifiers as theory-generating artifacts rather than unstructured text or human narratives. We employed several key components of the method, including open coding, memoing, axial coding, selective coding, and constant comparison, to analyze identifier names in the context of surrounding code and comments.

Our analysis reached theoretical saturation in that we developed axial codes iteratively until they accounted for all observed behaviors, refining them as new cases emerged. We also engaged in a form of core category development through cross-category synthesis, identifying broader trends that link behavioral interpretations across different grammatical constructs such as determiners, prepositions, digits, and conjunctions.

As grounded theory was originally developed in sociology, its application to the semi-structured nature of software artifacts, including lexical constructs like variable names, presents unique challenges. Some of these challenges have been explored in prior work on grounded theory in software engineering \cite{Adolph2011, Stol2016, hoda2022}, which informed our adapted approach. Our methodology is an example of how grounded-theory-based techniques can support the analysis of naming practices, particularly when grammar patterns and part-of-speech information are involved.

\subsection{Part of Speech Taggers}

POSSE \cite{Gupta:2013} and SWUM \cite{HillSWUM:2010}, and SCANL tagger \cite{newman_ensemble} are part-of-speech taggers created specifically to be run on software identifiers; they are trained to deal with the specialized context in which identifiers appear. Both POSSE and SWUM take advantage of static analysis to provide annotations. For example, they will look at the return type of a function to determine whether the word \textit{set} is a noun or a verb. Additionally, they are both aware of common naming structures in identifier names. For example, methods are more likely to contain a verb in certain positions within their name (e.g., at the beginning) \cite{Gupta:2013,HillSWUM:2010}. They leverage this information to help determine what POS to assign different words. Newman et al. \cite{Newman_GP} compared these taggers on a larger dataset than their original evaluation (1,335 identifiers) using five identifier categories: function, class, attribute, parameter, and declaration statement. They found that SWUM was the most accurate overall, with an average accuracy around 59.4\% at the identifier level. Later, Newman et al. created a new tagger that ensembled SWUM, POSSE, and Stanford together, then compared with SWUM, POSSE, and Stanford\cite{Toutanova:StanfordTagger} individually, finding that the ensembled tagger exceeded the others' performance metrics on identifiers \cite{newman_ensemble}.

\subsection{Human-subjects studies}
Several studies use human subjects to understand the influence and importance of different characteristics of identifiers. Our work is largely complementary to these studies, as it can be used in conjunction with data from these studies to create/support naming techniques. Reem et al. \cite{reem:2021} conducted a survey of 1100 professional developers, shedding light on developer preferences and practices regarding the content of identifier names, including the use of abbreviations and preferred identifier length. Feitelson et al. \cite{feitelson:2022} studied how the information content of identifiers named affected their memorability, and concluded that short names that contain focused information are likely optimal. Felienne et al. \cite{Felienne:2024} find, among other things, that while instructors agree on the importance of naming, there is disagreement between their teaching practices. Even internally, teachers are generally inconsistent in how they teach and practice identifier naming in the classroom. The results of their study highlight the importance of increasing our formal understanding of naming, which can help grow and support the consistency of teaching materials and practices in the classroom.

\subsection{Rename Analysis}
Arnoudova et al. \cite{Arnaoudova:2014} present an approach to analyze and classify identifier renamings. The authors show the impact of proper naming on minimizing software development effort and find that 68\% of developers think recommending identifier names would be useful. They also defined a catalog of linguistic anti-patterns \cite{Arnaoudova:2013}.  Liu et al.\cite{liu2015identifying} proposed an approach that recommends a batch of rename operations to code elements closely related to the rename. They also studied the relationship between argument and parameter names to detect naming anomalies and suggest renames \cite{liu2016nomen}. Peruma et al. \cite{Peruma:2018:EIW:3242163.3242169} studied how terms in an identifier change and contextualized these changes by analyzing commit messages using a topic modeler. They later extend this work to include refactorings \cite{Perumascam} and data type changes \cite{PERUMA2020110704} that co-occur with renames. Osumi et al. \cite{shinpei:2022} studied terms that were co-renamed with a goal of supporting developers in deciding when identifiers should be renamed together. In particular, they studied how location, data dependencies, type relationships, and inflections affected co-renaming.

These techniques are concerned with examining the structure and semantics of names as they evolve through renames. By contrast, we present the structure and semantics of names as they stand at a single point in the version history of a set of systems. Rename analysis and our work are complementary; our analysis of naming structure can be used to help improve how these techniques analyze changes between two versions of a name by examining changes in their grammar pattern. In particular, since we specifically study closed-category terms, rename analysis can leverage our results to improve its behavior on identifiers that contain these terms. For example, they might use our results to determine when to recommend a closed-category term during a rename operation.

\subsection{Identifier Type and Name Generation}
There are many recent approaches to appraising identifier names for variables, functions, and classes. Kashiwabara et al. \cite{Kashiwabara:2014} use association rule mining to identify verbs that might be good candidates for use in method names. Abebe \cite{Abebe:2013} uses an ontology that models the word relationships within a piece of software. Saeed et al. \cite{Saeed:2023} vectorize methods based on metrics and use the K-Nearst Neighbors algorithm with these vectors, and a large data set of methods, to recommend method names. Allamanis et al. \cite{Allamanis:2015} introduce a novel language model called the Subtoken Context Model. There has also been work to reverse engineer data types from identifiers \cite{malik2019,Hellendoorn:2018}. One thing these approaches have in common is the use of frequent tokens and source code context to try and generate high-quality identifier names (or understand their behavior for the purpose of generating types). There is a lot of work in this subfield, but the contrast to our work remains the same for all of them: These approaches aim to predict strong identifier names based on history. Our approach can help, since an understanding of common naming structures can support filtering out names that are inappropriate based on their grammatical structure; teach AI-based approaches how to optimize the identifiers they generate, or at least avoid using bad grammar structure; or help reverse-engineer the semantics of an identifier name based on its grammatical properties. In addition, automated name generation approaches cannot teach us much about naming practices on their own, nor can they help us formalize our understanding of strong naming structures and how those can be taught in a classroom. Thus, our work is novel, and complementary to identifier name generation approaches.

\subsection{Software Ontology Creation Using Identifier Names}
A lot of work has been done in the area of modeling domain knowledge and word relationships by leveraging identifiers \cite{AbebeDomainConcepts, RatiuProgramsKnowledgeBases, RatiuConceptMapping, Deissenboeck06aunified, falleriWordnet}. Abebe and Tonella \cite{AbebeDomainConcepts} analyze the effectiveness of information retrieval-based techniques for filtering domain concepts and relations from implementation details. They show that fully automated techniques based on keywords or topics have low performance but that a semi-automated approach can significantly improve results. Falleri et al., present a way to automatically construct a wordnet-like \cite{miller1995wordnet} identifier network from software. Their model is based on synonymy, hypernymy and hyponymy, which are types of relationships between words. Synonyms are words with similar or equivalent meaning; hyper/hyponyms are words which, relative to one another, have a broader or more narrow domain (e.g., dog is a hyponym of animal, animal is a hypernym of dog). Ratiu and Deissenboeck \cite{RatiuConceptMapping} present a framework for mapping real world concepts to program elements bi-directionally. They use a set of object-oriented properties (e.g., isA, hasA) to map relationships between program elements and string matching to map these elements to external concepts. This extends two prior works of theirs: one paper on a previous version of their metamodel \cite{Deissenboeck06aunified} and a second paper on linking programs to ontologies \cite{RatiuProgramsKnowledgeBases}. Many of these approaches need to split and analyze words found in an identifier in order to connect these identifiers to a model of program semantics (e.g., class hierarchies). All of these approaches rely on identifiers.

Many software word ontologies use meta-data about words to understand the relationship between different words. There is a synergistic relationship between the work presented here and software ontologies, as stronger ontologies can facilitate the effective generation and study of grammar patterns, and the CCID can aid in constructing stronger software word ontologies. In particular, studying closed-category terms helps strengthen the metadata used to generate an ontology that seeks to map how words are related to one another, or code behavior.

\subsection{Identifier Structure and Semantics Analysis}

Liblit et al.~\cite{Liblit06cognitiveperspectives} discuss naming in several programming languages and make observations about how natural language influences the use of words in these languages. Schankin et al. \cite{Schankin:2018} focus on investigating the impact of more informative identifiers on code comprehension. Their findings show the advantage of descriptive, compound identifiers over short single-word ones. Hofmeister et al \cite{Hofmeister:2017} compared comprehension of identifiers containing words against identifiers containing letters and/or abbreviations. Their results show that when identifiers contained only words instead of abbreviations or letters, developer comprehension speed increased by 19\% on average. Lawrie et al. \cite{Binkley2006} did a study and used three different ``levels'' of identifiers. The results show that full-word identifiers lead to the best comprehension compared to the other levels studied. Butler's work ~\cite{butler2010exploring} extends their previous work on Java class identifiers \cite{Butler:2009} to show that flawed method identifiers are also associated with low-quality code according to static analysis-based metrics. These papers primarily study the words found in identifiers and how they relate to code behavior or comprehension rather than word metadata (e.g., PoS).

Caprile and Tonella \cite{tonella:1999} analyze the syntax and semantics of function identifiers. They create classes which can be used to understand the behavior of a function; grouping function identifiers by leveraging the words within them to understand some of the semantics of those identifiers. While they do not identify particular grammar patterns, this study does identify grammatical elements in function identifiers, such as noun and verb, and discusses different roles that they play in expressing behavior both independently, and in conjunction, using the classes they propose. They also used the classes identified in this previous work to propose methods for restructuring program identifiers \cite{tonella:2000}. Fry and Shepherd \cite{Shepherd:2007,fry:2008} study verb-direct objects to link verbs to the natural-language-representation of the entity they act upon, in order to assist in locating action-oriented concerns. The primary concern in this work is identifying the entity (e.g., an object) which a verb is targeting (e.g., the action part of a method name). 

H{\o}st and {\O}stvold study method names as part of a line of work discussed in H{\o}st's dissertation \cite{hostdissertation}. This line of work starts by analyzing a corpus of Java method implementations to establish the meanings of verbs in method names based on method behavior, which they measure using a set of attributes which they define \cite{HostLexicon}. They automatically create a lexicon of verbs that are commonly used by developers and a way to compare verbs in this lexicon by analyzing their program semantics. They build on this work in \cite{hostphrasebook} by using full method names, which they refer to as phrases, and augment their semantic model by considering a richer set of attributes. The outcome is that they were able to aggregate methods by their phrases and come up with the semantics behind those phrases using their semantic model, therefore modeling the relationship between method names and method behavior. The phrases they discuss are similar to the general grammar patterns studied in our prior work \cite{Newman_GP}. They extend this use of phrases by presenting an approach to debug method names \cite{Host:2009}. In this work, they designed automated naming rules using method signature elements. They use the phrase refinement from their prior paper, which takes a sequence of PoS tags (i.e., phrases) and concretizes them by substituting real words. (e.g., the phrase $<$verb$>$-$<$adjective$>$ might refine to is-empty). They connect these patterns to different method behaviors and use this to determine when a method's name and implementation do not match. They consider this a naming bug. Finally, in \cite{Hst2011CanonicalMN}, H{\o}st and {\O}stvold analyzed how ambiguous verbs in method names makes comprehension of Java programs more difficult. They proposed a method to detect when two or more verbs are synonymous and used to describe the same behavior in a program, aiming to eliminate these redundancies while increasing naming consistency and correctness. They perform this detection using two metrics which they introduce, called nominal and semantic entropy. H{\o}st and {\O}stvold's work focuses heavily on method naming patterns; connecting these to the implementation of the method to both understand and critique method naming. 

Butler \cite{butler:2011} studied class identifier names and lexical inheritance, analyzing the effect that interfaces or inheritance has on the name of a given class. For example, a class may inherit from a super class or implement a particular interface. Sometimes this class will incorporate words from the interface name or inherited class in its name. His study builds on work by Singer and Kirkham \cite{singer:2008}, who identified a grammar pattern for class names of (adjective)* (noun)+ and studies how class names correlate with micro patterns. Among Butler's findings, he identifies a number of grammar patterns for class names: (noun)+, (adjective)+ (noun)+, (noun)+ (adjective)+ (noun)+, (verb) (noun)+ and extends these patterns to identify where inherited names and interface names appear in the pattern. The same author also studies Java field, argument, and variable naming structures \cite{butler:2015}. Among other results, they identify noun phrases as the most common pattern for field, argument, and variable names. Verb phrases are the second most common. Further, they discuss phrase structures for boolean variables; finding an increase in verb phrases compared to non-boolean variables. Olney \cite{Olney2016} compared taggers for accuracy on identifiers, but only on Java method names which were curated to remove ambiguous words (e.g., abbreviations).

Binkley et al \cite{Binkley:2011} studied grammar patterns for attribute names in classes. They come up with four rules for how to write attribute names: 1) Non-boolean field names should not contain a present tense verb, 2) field names should never only be a verb, 3) field names should never only be an adjective, and 4) boolean field names should contain a third-person form of the verb “to be” or the auxiliary verb “should”. Al Madi \cite{naser:2023} created a tool for performing lexical analysis of identifier names based on phonological, semantic, and orthographic similarity. Techniques that normalize identifiers, such as the one presented by Jingxuan \cite{Jingxuan:2023}, or by Hill \cite{hillamap} can help make generating grammar patterns easier by expanding abbreviations into full words that a tagger can recognize more accurately. Aman et al. \cite{Aman:2024} studied confusing variable pairs, which are variables with very similar names, to understand how/if they are changed over time, and how pervasive they are.

None of the projects in this subsection deal specifically with closed-category grammar patterns, or even terms that fall within a closed PoS category. Many of them, particularly the work on PoS taggers, on grammar patterns in differing contexts, normalizing identifier names, and on grammatical anti-patterns, are likely mutually-synergistic to our work. This is because a stronger understanding of closed-category terms/patterns, and how they relate to program behavior, can help support the style of analysis these works leverage.

\section{Discussion and Future Work} \label{discussion}

Closed-category terms—including determiners, prepositions, conjunctions, and numerals—are compact grammatical structures that carry dense behavioral meaning in source code. Developers use them to embed logical cues, control flow, and intent directly into identifiers. Based on our findings, we outline several actionable implications for tool builders, educators, and developers:

\begin{enumerate}
    \item \textbf{Treat closed-category terms as lightweight cognitive annotations.}
    Developers should consider how they might improve identifier clarity by deliberately using closed-category terms to signal concepts such as selection (\texttt{allItems}), temporality (\texttt{nextNode}), negation (\texttt{noCache}), disjunction (\texttt{keyOrIV}), or encode system-specific roles/concepts (e.g., \texttt{arg1}, \texttt{Neo4j}). These terms encode behavioral roles that may be missed in names relying solely on open-class words. \emph{Naming checkers, LLMs, or documentation generators should highlight or recommend these roles to aid consistency and reader comprehension, as well as assist developers in reflecting on how they might improve their terminology usage in identifiers. Our axial codes offer a schema that could seed such tools.}

    \item \textbf{Augment use of closed-category terms by considering code context.} Our data show that closed-category term usage varies systematically across code contexts. For example, determiners like \texttt{no}, \texttt{next}, and \texttt{this} are very unlikely to appear in class names, while prepositions such as \texttt{on} or \texttt{to} are common in function names but rare elsewhere. \emph{As developers consider when to use closed category terms in their coding, they can leverage context to help in their decision making. Tool support can make this easier-- IDEs and linters should offer context-aware naming prompts, e.g., flagging uncommon use of conjunctions in class names or suggesting appropriate determiners for declarations.}

    \item \textbf{Use grammar patterns to reveal relational structure.}
    Closed-category terms not only express behavior, they also help structure it. Patterns like \texttt{DT NM N} and \texttt{P NM N} typically signal unary relations (e.g., \texttt{lastError}, \texttt{asFloat}), while \texttt{N CJ N} or \texttt{N P N} often signal binary ones (e.g., \texttt{dataOrLogger}, \texttt{readFromDisk}). Function names with final prepositions (e.g., \texttt{sendTo}) sometimes encode a second operand passed via parameters. \emph{Developers should consider the latent logical structure of the grammatical patterns they use when naming. Static analyzers or naming tools could infer missing operand roles or flag relational ambiguity. For example, encountering a binary-looking pattern in a context that provides only one operand could trigger a naming prompt.}
\end{enumerate}

These findings have practical implications for tool builders and educators interested in improving naming support. For example, grammar patterns could be integrated into static analysis or IDE plugins to provide optional, contextual suggestions; highlighting when an identifier follows an uncommon structure or when the pattern contrasts with its code context. This does not imply the name is incorrect, but it may prompt a developer or reviewer to reflect on whether the chosen pattern aligns with the intended semantics. For instance, encountering a pattern like \texttt{N CJ N} (e.g., \texttt{dataOrLogger}) in a constant declaration could trigger a soft prompt: “This naming pattern is rare in this context—consider whether it clearly communicates its role.”

Grammar patterns can also be used to scaffold naming suggestions in LLM-driven tools. Instead of generating identifiers purely from task descriptions, models could be prompted to produce names that instantiate common open- or closed-category structures (e.g., \texttt{DT NM N}, \texttt{P N}), which are frequently associated with behavioral semantics. This approach may help align completions with human naming conventions, while still allowing flexibility in term choice.

Beyond tooling, grammar pattern awareness can enhance educational workflows. Instructors could use pattern frequency and semantics to illustrate naming “idioms,” helping students understand how experienced developers encode behavior through compact syntactic forms. Similarly, code review tools might use grammar pattern summaries to draw attention to unconventional naming constructs, offering reviewers an additional signal without enforcing rigid standards.

In future work, we plan to evaluate these ideas empirically: measuring whether adherence to common patterns improves comprehension, how grammar scaffolding affects naming quality in LLM-generated identifiers, and whether tools that surface grammar patterns can meaningfully assist developers. In addition, it would be interesting to perform studies simialar to Schankin \cite{Schankin:2018} and Hofmeister \cite{Hofmeister:2017}'s work on how the descriptiveness of names influence comprehension; or Arnaoudova \cite{Arnaoudova:2013}, and Host\cite{Host:2009} who looked at how code behavior and naming structure could be used to measure name quality. Specifically, we could study how closed-category terms change or augment the outcomes of their studies.

\section{Threats to Validity}\label{threats}

\textbf{Construct Validity}: This study is conducted on a manually annotated dataset of 1,275 identifiers containing closed-category terms, the largest of its kind at the time of writing. A potential threat lies in the completeness of our closed-category term list: we relied on a predefined lexicon (Section~\ref{methodology}), meaning novel or unlisted terms may be absent from the dataset. However, their absence would likely expand our results rather than refute them. Our identifier sample is restricted to production code, and although we excluded known test files, developers may occasionally include test logic in production files. To mitigate this, we manually reviewed each identifier and its source context. Furthermore, while we used file extensions to distinguish C (.c, .h) from C++ (.cpp, .hpp), these conventions are not absolute. We addressed this by manually validating the source language of each identifier.

\textbf{Internal Validity}: Abbreviations within identifiers were not expanded, which may have caused occasional misinterpretation by annotators. However, annotators had access to the surrounding source code, reducing the risk of misannotation. Grammar pattern tagging and axial coding for each closed-category term were both subject to cross-validation by three independent annotators and evaluated using Fleiss’ Kappa to assess agreement. We used a grounded-theory-inspired approach to develop our behavioral codes. Four coders participated in open and axial coding; during the selective coding phase, one coder proposed all selective codes, which were then validated and refined collaboratively by the other three through discussion until thematic saturation was reached.

We used statistical methods to examine correlations between closed-category terms and contextual variables. We performed two chi-square tests: one to assess correlation between closed-category part-of-speech categories (e.g., Determiner, Preposition) and programming language (Java, C, C++), and another for their correlation with code context (Attribute, Function, Declaration, Parameter, Class), derived automatically via srcML \cite{collard:2016}. We applied Bonferroni correction to account for multiple comparisons. A threat to internal validity is the assumption of independence in the chi-squared test. If violated, some significant values may be distorted. However, the primary insights of RQ1, which focus on behavioral coding through qualitative analysis, are unaffected by this statistical assumption.

\textbf{External Validity}: Our data includes identifiers from C, C++, and Java, three widely used languages with similar syntactic and object-oriented paradigms. While this helps reduce language-specific bias, our findings may not generalize to other paradigms such as functional or logic-based languages, where naming conventions and code contexts may differ significantly.

\textbf{Mitigation Strategies}: To ensure transparency and reproducibility, the dataset will be made publicly available (Section~\ref{data_avail}). Annotators were allowed to inspect source code when labeling identifiers, and each identifier was independently annotated twice. Grammar patterns and axial codes were validated by multiple annotators, with inter-rater agreement assessed using Fleiss’ Kappa. We selected a representative sample from 30 software systems, sized to meet a 95\% confidence level with a 5\% confidence interval. Code context was derived automatically using \texttt{srcML}. Finally, to evaluate whether closed-category term usage varies by domain, we curated a domain-specific dataset (e.g., compilers, databases, networking tools) and compared it against a general-purpose set selected without regard to domain. We applied a Mann-Whitney U test to compare term frequencies between these groups, normalizing by lines of code to control for system size.

\section{Conclusions}\label{conclusions}

This paper presents a detailed empirical study of \textbf{closed-category terms in identifier names}, highlighting how they function semantically across a wide range of software artifacts. Our contributions include:

\begin{enumerate}
    \item \textbf{The CCID dataset:} A new, part-of-speech–annotated dataset of identifier names containing closed-category terms, released with this paper. It supplements prior datasets focused on open-class lexical items, enabling more nuanced research into naming semantics.
    
    \item \textbf{A dual-level coding framework:} We introduce a combined selective and axial coding scheme to interpret the semantics of determiners, prepositions, conjunctions, and numerals in identifiers. This framework maps grammar structure to conceptual behavior in a way not previously formalized, and provides a solid basis for future research and development of naming practices.
    
    \item \textbf{Insights into the distinct semantic roles of closed-category words:} Our study shows how programming diverges from natural language in its use of terms like \texttt{next}, \texttt{both}, and \texttt{or}, and how these terms reflect functional intent embedded in code structure.
\end{enumerate}

These findings have practical implications for code review, refactoring, and intelligent naming support. Automated tools can leverage our work to suggest context-appropriate names, detect inconsistent patterns, or provide just-in-time feedback during the development process. The insights we have uncovered can be packaged and curated by educators to teach semantic clarity in naming, moving beyond generic best practices to behavior-specific guidance.

We also outline several promising directions for future research:

\begin{enumerate}
    \item Further validation of the behavioral categories we propose, especially through comprehension studies that measure how different naming structures affect developer understanding. This could also extend to examining the correlation between the beahvoral categories and software quality metrics.
    
    \item Experimental work to test which closed-category naming patterns improve or hinder program comprehension, while controlling for code context and developer experience.
    
    \item Integration of our findings into intelligent development tools, including naming recommendation systems, refactoring support, and automated code reviewers capable of detecting semantic mismatches or naming anti-patterns.
\end{enumerate}

This work demonstrates that closed-category terms are not linguistic noise; they are deliberate, behaviorally meaningful tools in the software naming arsenal. By mapping their roles across grammar patterns and program contexts, we provide a foundation for future studies in naming semantics and for tools that support naming literacy. We believe our findings raise the question: \textbf{Should closed-category terms be used more often in naming?} It is clear that they are used purposefully, and we believe that their use does help comprehension, but in which situations is that true? And how can we detect when such terms should be used, as opposed to another naming pattern? We believe these findings can help guide tool builders and researchers in novel, fruitful directions that formalize and improve naming practices among developers.

\section{Declarations}
\subsection{Funding} This work was not funded by any agency
\subsection{Ethical approval} This work does not involve human or animal subjects and does not require IRB approval
\subsection{Informed consent} This work does not involve human or animal subjects and does not require informed consent
\subsection{Author Contributions}
\begin{enumerate}
    \item Conceptualization: Christian D. Newman

\item  Data curation: Christian D. Newman, Anthony Peruma, Syreen Banabilah, Michael J. Decker, Reem S. AlSuhaibani, Eman Abdullah Alomar, Mahie Crabbe

\item  Formal analysis: Christian D. Newman

\item  Investigation: Christian D. Newman

\item  Methodology: Christian D. Newman

\item Project administration: Christian D. Newman

\item  Resources: Christian D. Newman

\item  Software: Christian D. Newman, Farhad Akhbardeh, Anthony Peruma

\item  Supervision: Christian D. Newman

\item Validation: Christian D. Newman, Anthony Peruma, Syreen Banabilah, Eman Abdullah Alomar

\item Visualization: Christian D. Newman

\item Writing – original draft: Christian D. Newman, Jonathan I Maletic, Mohamed Wiem Mkaouer, Marcos Zampieri

\item Writing – review \& editing: Christian D. Newman, Jonathan I Maletic, Anthony Peruma, Syreen Banabilah, Michael J. Decker, Reem S. AlSuhaibani, Eman Abdullah Alomar, Marcos Zampieri
\end{enumerate}

\subsection{Data Availability Statement }\label{data_avail}
We have created a repository that contains the data and scripts needed to generate the numbers and statistical analysis from the RQs. The scripts are in the scripts directory and need only to be run (they take no arguments). The data directory contains all annotation data broken down by closed category. Each file is named after the category it contains, and they are amalgamated in a single file (called Tagger Open Coding). This repository can be found at this link\footnote{https://github.com/SCANL/closed\_category\_emse\_analysis\_scripts}.

\subsection{Conflict of Interests}
We have no competing/conflicting interests to report

\subsection{Clinical Trial Number} Not Applicable


\newpage
\bibliography{main}
\end{document}